\documentclass{aa}

\usepackage{graphicx}
\usepackage{txfonts}

\usepackage{lscape}

\usepackage[colorlinks=true, citecolor=blue, linkcolor=blue, urlcolor=blue]{hyperref}
\usepackage[dvipsnames]{xcolor}
\AtBeginDocument{\let\linenumbers\relax \let\resetlinenumber\relax \nolinenumbers}
\begin{document}
\nolinenumbers
%%%%%%%%%%%%%%%%%%%%%%%%%%%%%%%%%%%%%%%%

%%%%%%%%%%%%%%%%%%%%%%%%%%%%%%%%%%%%%%%%
   \title{Astrometric modeling of unresolved variable binary systems}
   \subtitle{I. Application to Gaia DR4 epoch astrometry of classical Cepheids}

%%%%%%%%%%%%%%%%%%%%%%%%%%%%%%%%%%%%%%%%

%%%%%%%%%%%%%%%%%%%%%%%%%%%%%%%%%%%%%%%%

   \author{K. Sivkova\inst{1,2}\corrauth{ekaterina.sivkova@obspm.fr}
        \and P. Kervella\inst{1,2} \email{Pierre.Kervella@obspm.fr}
        \and A. Gallenne\inst{3} \email{agallenne@academicos.uta.cl}
        \and L. Decin\inst{4} \email{leen.decin@kuleuven.be}
        \and N. Nardetto\inst{5} \email{Nicolas.Nardetto@oca.eu}
        \and F. Arenou\inst{6,1} \email{Frederic.Arenou@obspm.fr}
        \and A. Mérand\inst{7} \email{amerand@eso.org}
        \and N.R. Evans\inst{8} \email{nevans@cfa.harvard.edu}
        \and B. Apostolova\inst{5} \email{brankica.apostolova@oca.eu}
        \and M. C. Bailleul\inst{5} \email{manon.bailleul@oca.eu}
        \and R. S. Rathour\inst{5} \email{rajeev.rathour@oca.eu}
        \and W. Kiviaho\inst{1,2} \email{Wilma.KIVIAHO@obspm.fr}
        \and G. Bras\inst{1} \email{garance.bras@obspm.fr}
        \and V. Hocdé\inst{5} \email{vincent.hocde@oca.eu}
        \and L. Breuval\inst{9} \email{lbreuval@stsci.edu}
        \and C. Vert\inst{5,10} \email{cedric.vert@oca.eu}
        \and D. P. González\inst{1,2} \email{daniela.gonzalez@obspm.fr}
        }

   \institute{LIRA, Observatoire de Paris, Université PSL, Sorbonne Université, Université Paris Cité, CY Cergy Paris Université, CNRS, 5
    place Jules Janssen, 92195 Meudon, France
    \and French-Chilean Laboratory for Astronomy, IRL 3386, CNRS and U. de Chile, Casilla 36-D, Santiago, Chile
    \and Instituto de Alta Investigación, Universidad de Tarapacá, Casilla 7D, Arica, Chile
    \and Institute of Astronomy, KU Leuven, Celestijnenlaan 200D, 3001, Leuven, Belgium
    \and Université Côte d’Azur, Observatoire de la Côte d’Azur, CNRS, Laboratoire Lagrange, Nice, France
    \and UNIDIA, Observatoire de Paris, Université PSL, CNRS, 92190 Meudon, France
    \and European Southern Observatory Headquarters, Karl-Schwarzschild-Str. 2, 85748 Garching, Germany
    \and Smithsonian Astrophysical Observatory, MS 4, 60 Garden Street, Cambridge, MA 02138, USA
    \and European Space Agency (ESA), ESA Office, Space Telescope Science Institute, 3700 San Martin Drive, Baltimore, MD 21218, USA
    \and MAUCA – Master track in Astrophysics, Université Côte d’Azur \& Observatoire de la Côte d’Azur, Parc Valrose,
    06100 Nice, France
   }

   \date{Received 12 June 2026, Accepted 4 November 2026}

  \abstract
   {The fourth data release of the \textit{Gaia} mission (DR4) will, for the first time, provide epoch astrometry. When analysing these data, binary sources must be treated carefully, as modelling their astrometry as single stars can bias parallax and proper motion estimates.}
  {We aim to develop a framework to simulate and interpret epoch astrometry of unresolved binary systems and to evaluate the impact of model assumptions on the resulting astrometric parameters for the systems considered here.}
   {We simulate \textit{Gaia} DR4-like epoch astrometry for a sample of 17 well-characterised unresolved binary systems, including both non-pulsating binaries and systems containing one photometric variable component -- so-called variability-induced movers (VIMs).}
   {We present the new \texttt{vimbinary} tool used to simulate and analyse the epoch astrometry data. We show that both binary motion and the VIM effect can significantly impact parallax and proper motion measurements, with the magnitude of the effect strongly dependent on the orbital period, photocentric amplitude, and flux ratio of the specific system. We also show that larger time coverage does not necessarily improve parallax precision, in contrast to proper motion, for which precision generally improves with longer baselines. Instead, parallaxes retrieved with binary-system fitting are more accurate than those obtained within the single-star framework.}
   {We presented simulations of Gaia epoch astrometry, which is not available at the time of writing, that allows us to prepare future DR4 analyses. Our framework is applicable to VIM binaries with complex astrometric behaviours, in particular classical Cepheids that are generally found in multiple systems.}


   \maketitle
\nolinenumbers

\section{Introduction}~\label{Sec:Intro}

The Gaia third data release \citep[DR3,][]{Gaia, DR3} provided 800\,000 binary solutions \citep{Arenou_2023}, which represents a significant increase in the total number of known binaries. Nevertheless, the whole catalogue contains about 1.5 billion sources with at least a five-parameter (single-star) solution; given the expected binary fraction of at least 50\% for Sun-like stars, rising to nearly 100\% for the most massive stars \citep{Moe2017ApJS..230...15M, Fulton2018AJ....156..264F, Offner2023ASPC..534..275O}, the majority of binary systems likely remain undetected. This misrepresentation affects the quality of the solutions and the accuracy of the retrieved parameters.
  
The upcoming DR4 is expected to become groundbreaking for binary searches. Not only is it expected to contain a drastically increased number of binary solutions compared to DR3, but the community will also have access to epoch astrometry for the first time. This will allow users to conduct more targeted studies and to use complementary data, moving beyond the binary detection performed in DR3 \citep{Halbwachs_2023}, which applied sharp cuts to avoid suspicious solutions. Here, we focus on the treatment of binary systems in epoch astrometry in preparation for the DR4.
  
\begin{figure}[hbt!]
  \centering \includegraphics[width=.99\hsize]{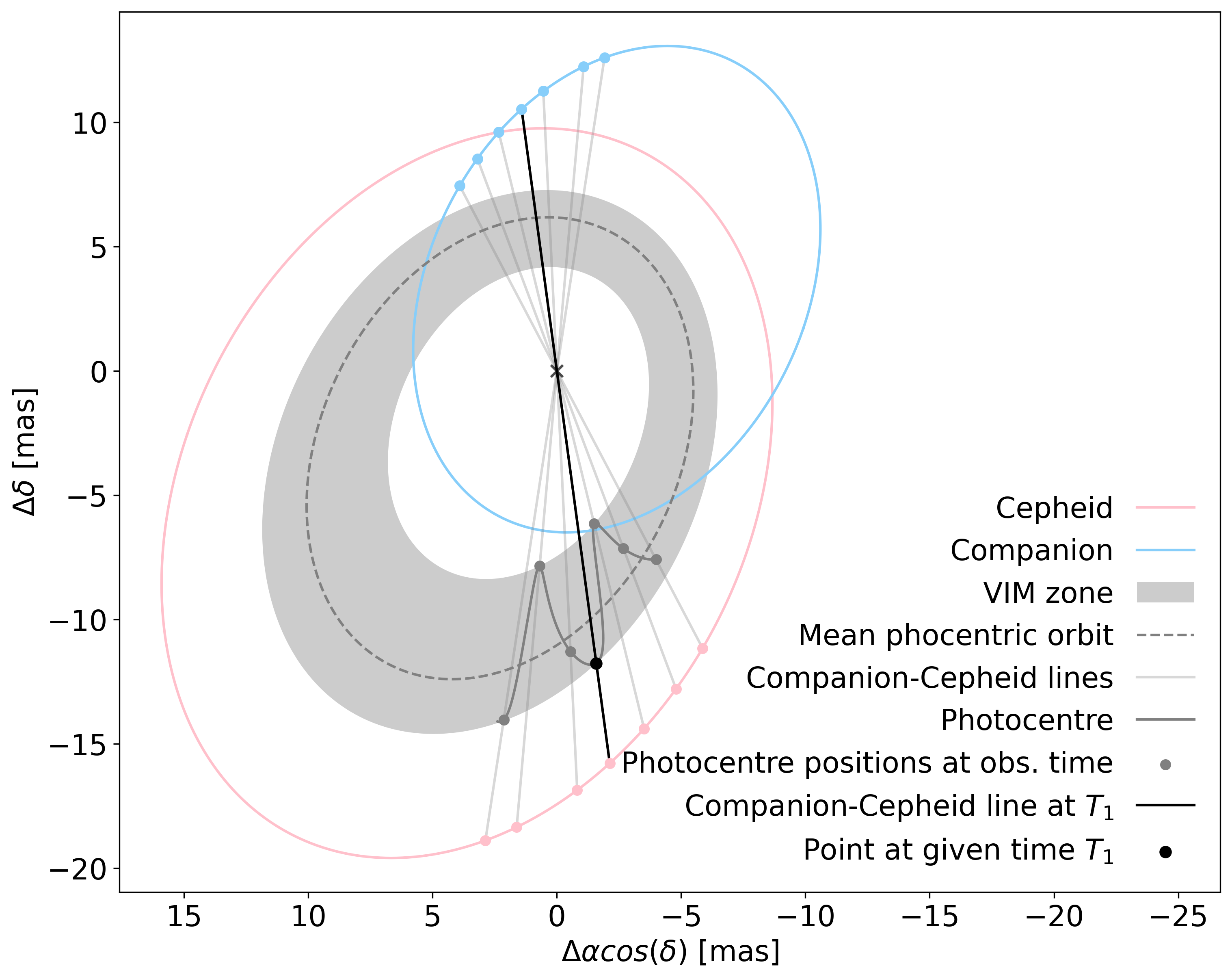}
  \caption{Conceptual illustration of the VIM effect, with the effect exaggerated for clarity. The two components' orbits (in colours) centred on the centre of mass (grey cross), with the corresponding mean photocentric orbit (dashed line). Variability shifts the photocentre along the line connecting the two components. These shifts are bounded within the VIM zone (grey area), corresponding to the minimum and maximum light of the Cepheid. At a given time, Gaia observes a single photocentric position determined by both the orbital phase and the pulsation phase of the Cepheid.}
  \label{VIM_exp} 
\end{figure}

A specific type of binary systemin detectable with Gaia astrometry are the so-called \textit{variability-induced movers} (VIMs). These systems are defined as unresolved binary systems containing one photometrically variable component. As the flux of the variable component changes, its contribution to the total flux varies, which shifts the position of the photocentre back and forth between the two components (see Fig.~\ref{VIM_exp}). This effect was first studied by \citet{Wielen_1996} in the context of the \textsc{Hipparcos} mission, with possible extension to Gaia. Such missions provide an efficient way to detect these systems because they require a large number of highly precise astrometric measurements over a long time span.

Cepheids are prime VIM candidates because they exhibit a high fraction of binary or multiple systems \citep{Evans_2015, Sana_2016, Kervella_2019a}, and their pulsational variability is continuous and stable over time, making them particularly suitable for this method \citep{Wielen_1996}. Moreover, observations of Cepheids with this approach are especially important as Cepheids are a cornerstone of the cosmic distance ladder. Their period--luminosity (PL) relation \citep{Leavitt_1908, Leavitt_1912} provides one of the primary routes to determine the Hubble constant ($H_0$) \citep{Riess_2022, Breuval_2024, H0DN}. In the Gaia DR4 era, we expect to obtain precise parallaxes for a large sample of Cepheids to calibrate this relation. In addition, Cepheid masses suffer from discrepancies between values derived from evolutionary models and those inferred from pulsation calculations \citep{Neilson_2011}. Binary systems provide a way to measure Cepheid masses dynamically, helping to calibrate and constrain these models.

In Gaia DR3, only two Cepheids were treated as binary systems in the astrometric solutions \citep{Arenou_2023, DR3}. One of them, Gaia DR3 5256789350372915328, previously unknown as a binary in the literature, was modelled using a constant-acceleration solution, in which the source motion is approximated by a linear change in proper motion over time. The second, RX~Cam, already suspected to host a companion \citep{Szabados_1992}, was fitted with a full Keplerian solution. However, as noted by \citet{Shetye_2024}, the derived orbital period coincides with the 33-month time span of DR3, which raises concerns about the robustness of the solution. Nevertheless, \citet{Groenewegen_2013} derived an orbital period of 1113.8~d by using radial velocities, comparable to the 33-month duration of DR3. The validity of the RX~Cam orbital solution therefore remains an open question. No Cepheids were treated as VIMs in DR3, likely reflecting the limited number of VIM solutions in the catalogue (869 in total) and the fact that Cepheids are generally distant objects, so their apparent orbital motion often remains small.

As a result, comparisons of parallaxes for binary Cepheids between Gaia DR3 values and those derived from the orbital method, which combines relative astrometry and radial velocity measurements, reveal discrepancies \citep{Gallenne_2019, Evans_2025, Gallenne_2025}. In some cases, these differences reach several standard deviations, such as for V1334~Cyg, where a $3.6\sigma$ offset was reported with respect to DR2 \citep{Gallenne_2018} and increases to $4.1\sigma$ in DR3. In such systems, binarity and pulsation can both contribute to the parallax differences. These physical effects influence the Gaia observables and, if not properly modelled, introduce biases in the inferred parameters. As we demonstrate in this work, both parallax and proper motion can be affected.
  
We expect this situation to improve with the upcoming DR4 release \citep[expected on December 2nd of 2026,][]{Brown_2025}, which will provide access to epoch astrometry. As these data are not publicly available at the time of writing, in this work, we introduce the \texttt{vimbinary} tool to simulate Gaia DR4-like epoch astrometry, with particular emphasis on VIM effects. The simulated data are then fitted and analysed under the single-star assumption commonly adopted in Gaia DR3. This allows us to quantify the impact of binarity and variability-induced motion in case of the considered systems on the inferred astrometric parameters, in particular parallax and proper motion.

Section~\ref{Method} describes the general principles of Gaia astrometry and the modelling formalism adopted for the different models. Section~\ref{Validation} compares our simulations with existing datasets and solutions, thereby validating our implementation. Section~\ref{Sec:sample} describes the sample of binary systems considered in this work. Section~\ref{Sec:analysis} discusses the impact of VIM and binarity on proper motion and parallax for the systems considered here, while Section~\ref{Conclusion} presents our conclusions.

\section{Modelling Gaia epoch astrometry \label{Method}}
   
The following sections introduce the general principles of Gaia epoch astrometry and describe their implementation in our dedicated Python package \texttt{vimbinary}\footnote{\texttt{vimbinary}: \href{https://doi.org/10.5281/zenodo.22118250}{https://doi.org/10.5281/zenodo.22118250}}.

\subsection{Gaia along-scan astrometry \label{Sec:AL_astro}}

\begin{figure*}[hbt!]
  \centering \includegraphics[width=.99\hsize]{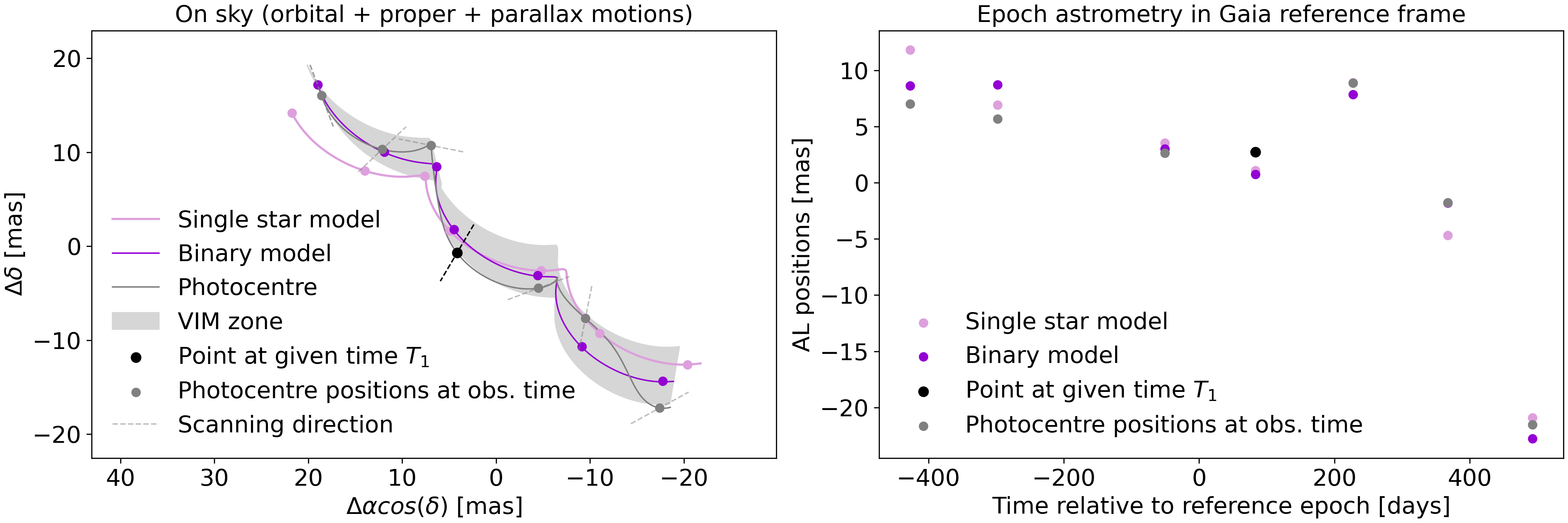}
  \caption{The same VIM system as in Fig.~\ref{VIM_exp} with the VIM effect exaggerated, showing the astrometric position of the object for different assumptions about its nature: VIM, binary, and single star. Left: sky-plane astrometric motion, including the combined effects of orbital, proper motion, and parallax motions. The dashed lines indicate the direction of the along-scan at each observation epoch. Right: the corresponding astrometric positions projected onto the Gaia along-scan reference frame.}
   \label{vim_exp2}
\end{figure*}

The concept of along-scan (AL) measurements follows directly from the Gaia observing strategy. Rather than measuring positions in right ascension ($\alpha$) and declination ($\delta$) at each epoch, the satellite continuously scans the sky and records a time series of measurements in its own reference frame: the along-scan abscissa $w_{\mathrm{AL}}$ (hereafter simply $w$) and the across-scan coordinate $w_{\mathrm{AC}}$. Only the along-scan coordinate is determined with high precision, effectively making each observation one-dimensional \citep{Lindegren_2012}. See Fig.~\ref{vim_exp2} for a visual example.

These measurements, referred to as epoch astrometry, are time-dependent. They consist of a time series of along-scan positions with respect to the reference position adopted in the global astrometric solution. Each measurement is associated with a scan angle and a parallax factor, resulting in a set of three time-dependent quantities.

The scan angle is the position angle of the along-scan direction on the sky, measured from local North towards East. It varies with the orientation of the \textit{Gaia} spacecraft at the time of the observation. The parallax factor is the projection coefficient that converts the source parallax into the corresponding along-scan displacement. It depends on the relative geometry of the source, the Sun, and the spacecraft at the observing epoch.

Because each scan provides only a one-dimensional constraint, we cannot directly represent the source motion in two dimensions. However, the large number of scans obtained at different position angles lifts this degeneracy, allowing the full two-dimensional motion of the source to be recovered.

As mentioned above, the observation epochs, scan angles, and parallax factors are determined by the Gaia scanning law and the sky coordinates of the source. We retrieve them from the Gaia observation forecast tool GOST\footnote{GOST: \href{https://gaia.esac.esa.int/gost/}{https://gaia.esac.esa.int/gost/}}. This tool is based on Gaia scanning law, but it does not take into account operational activities. In practice, about 20 percent of the data points are lost because of outlier rejection, dead time, and other effects. We accounted for this loss by randomly rejecting 10\% of all predicted transits together with an additional 10\% of per-CCD points, consistent with real time-sampling.
   
Gaussian Gaia-like measurement noise was then added to the simulated epoch astrometry using the magnitude-dependent empirical CCD along-scan uncertainty prescription described by \citet{ElBadry2024OJAp....7E.100E}, based on the median DR3 AL uncertainties \citep[see also][]{2023A&A...674A..10H}, which depend on the system G magnitude. The individual AL measurements are assumed to be independent and uncorrelated, following the same simplifying assumption adopted by the astrometric global iterative solution \citep[AGIS,][]{Lindegren_2012}, as well as by \citet{Everall_2021} and \citet{ElBadry2024OJAp....7E.100E}.

\subsection{Single-star astrometric signal simulation} \label{Sec:single_star_forward}

The epoch astrometry of a single star (SS) represents the change in position due to proper motion (PM) and parallax, projected onto the Gaia along-scan direction:
\begin{equation}
w_{\mathrm{ss}} = (\Delta \alpha^\star + \mu_{\alpha^\star} t) \sin \theta + (\Delta \delta + \mu_{\delta} t) \cos \theta + \Pi_{\varpi} \varpi,
\label{Eq:SS}
\end{equation}
where $\Delta \alpha^\star$ and $\Delta \delta$ denote the offsets from the reference position $(\alpha_0, \delta_0)$, $\mu_{\alpha^\star}$ and $\mu_{\delta}$ are the proper motion components, $\varpi$ is the parallax, and $\Pi_{\varpi}$ the corresponding along-scan parallax factors. All quantities are projected onto the Gaia reference frame using the scan angle, $\theta$. The time $t=t_{\mathrm{obs}}-t_{\mathrm{ref}}$ is relative to a reference epoch $t_{\rm ref}$, which is defined for each data release ($t_{\rm ref}=2017.5$ for DR4).

Eq.~\ref{Eq:SS} is linear in the astrometric parameters ($\Delta \alpha^\star$, $\Delta \delta$, $\mu_{\alpha^\star}$, $\mu_{\delta}$, $\varpi$), allowing for a straightforward weighted least-squares solution (see App.~\ref{annex:fitting}). This linearity is consistent with the approach adopted in Gaia DR3, where astrometric parameters are estimated within a linearised least-squares framework. While this ensures robustness and computational efficiency, any unmodelled effects, such as orbital motion or variability-induced photocentre shifts, can be partially absorbed into the fitted parameters, leading to biased estimates.

\subsection{Binary astrometric signal simulation} \label{Sec:binary_simulation}

The retrieval of binary astrometric signals can be performed at different levels of complexity, depending on the orbit coverage provided by the data. In the context of the Gaia DR3 non-single star (NSS) processing, \citet{Halbwachs_2023} considered a hierarchy of models for non-pulsating systems, including constant-acceleration, variable-acceleration, and full orbital solutions. The choice of model is driven by how well the orbital motion is detected and sampled over the observing time span.

A similar hierarchy exists for variability-induced movers. \citet{Wielen_1996} introduced a sequence of models of increasing complexity: fixed (VIMF), linear (VIML), acceleration (VIMA), and orbital (VIMO). The simpler models approximate the photocentre motion when the orbital signal is only partially resolved, while the VIMO model provides a full Keplerian description of the photocentre displacement. The fixed model, which assumes a static geometry between the components, was already implemented in Gaia DR3. The explicit expressions for the VIMF, VIML, VIMA, and VIMO models are provided in App.~\ref{annex:vim}.

In general, the along-scan signal of a binary system (BS) can be written as
\begin{equation}
\begin{aligned}
\label{w_bs}
w_{\mathrm{bs}} &= (\Delta \alpha^\star + \mu_{\alpha^\star} t + \alpha^\star_{\mathrm{orb}}(t)) \sin \theta \\
       &+ (\Delta \delta + \mu_{\delta} t + \delta_{\mathrm{orb}}(t)) \cos \theta \\
       &+ \Pi_{\varpi} \varpi \\
       &= w_{\mathrm{ss}} + \alpha^\star_{\mathrm{orb}}(t) \sin \theta + \delta_{\mathrm{orb}}(t) \cos \theta,
\end{aligned}
\end{equation}
where $\alpha^\star_{\mathrm{orb}}(t)$ and $\delta_{\mathrm{orb}}(t)$ describe the additional photocentre motion induced by binarity and, in the case of VIMs, by variability, while the $w_{\mathrm{ss}}$ term describes the motion of the centre of mass. In this work, we adopt the full orbital description for our forward simulations of Gaia epoch astrometry.

For an unresolved binary system, Gaia measures the motion of its photocentre. We therefore model the orbital contributions $\alpha^\star_{\mathrm{orb}}(t)$ and $\delta_{\mathrm{orb}}(t)$ through the photocentric orbit. This orbit can be expressed as a scaled version of the true relative orbit, with a semi-major axis given by
\begin{equation}
a_{\mathrm{ph}} = a_{\mathrm{rel}} \, |B - \beta|,
\label{Eq:aph}
\end{equation}
where $a_{\mathrm{rel}}$ is the (generally unknown) semi-major axis of the relative orbit of the companion (star~2) with respect to the primary (star~1), $B = M_2/(M_1 + M_2)$ denotes the mass ratio, and $\beta = F_2/(F_1 + F_2) = F_2/F$ is the flux ratio, with $F$ the total system flux. We then compute the photocentric motion for a non-variable system using a Keplerian orbit of the binary system with the derived $a_{\mathrm{ph}}$.

\subsection{Variability-induced motion simulation} \label{Sec:vim_simulation}

In the case of VIMs, the variability of $F_1(t)$ introduces a time dependence in the total flux. As a result, the photocentre position, which already varies because of the orbital motion, also becomes dependent on the stellar flux variations and can be computed as
\begin{eqnarray}  
x_{\mathrm{ph}} (t) & = & \frac{F_1(t) x_1(t) + F_2 x_2(t)}{F_1(t) + F_2} \nonumber \\
& = & (1-\beta(t)) x_1(t) + \beta(t) x_2(t) 
\label{Eq:x_ph}
\end{eqnarray}
where $x_i(t)$ denotes the position of one component in $\alpha^\star$ or $\delta$ along its orbit around the centre of mass.  This yields the sky-projected photocentre coordinates $\alpha^\star_{\mathrm{orb}}(t)$ and $\delta_{\mathrm{orb}}(t)$, which we then use in Eq.~\ref{w_bs}.

Thus, the flux variations translate directly into changes in the photocentre position, and hence photometric and astrometric data can be fitted simultaneously to constrain the variability-induced motion (see App.~\ref{annex:vim}). We note that it is not necessary to constrain the light curve itself; only the factor relating the flux variations to the photocentre displacement needs to be determined. In the case where both components are variable, the photocentre positions will reflect the combination of both variations, resulting in a more complex situation. Given that most Cepheids have photometrically stable companions, we do not consider this case in this work.

Simulating a VIM system therefore requires modelling the stellar flux variation. For Cepheids, we recover their photometric variation from their Fourier parameters derived in DR3 \citep{Eyer_2023} in the catalogue \citep[\texttt{gaiadr3.vari\_cepheid},][]{DR3var}.
   
To quantify the contribution of the VIM effect, it is useful to define the amplitude of the photocentre displacement induced by the flux variability. Following \citet{Wielen_1996}, we define
\begin{equation}
d_{\mathrm{var}}(t) = F_2 d_{12}(t) \left( \frac{1}{F_{\min}} - \frac{1}{F_{\max}} \right)\,,
\label{Eq:d_var}
\end{equation}
where $F_{\min}$ and $F_{\max}$ are the minimum and maximum values of the variable total flux $F(t) = F_1(t) + F_2$, and $d_{12}(t)$ is the separation between the two components. This quantity defines the width of the ''VIM zone'' (see Fig.~\ref{VIM_exp}), a band around the mean photocentric orbit that delimits the maximum and minimum photocentric shifts introduced by the VIM effect. The width of this band varies along the orbit because it depends on the separation between the two stars, reaching its maximum when $d_{12}=a_{\mathrm{rel}}$.

\subsection{Correction of proper motion anomaly for unresolved binaries \label{correction}}

Almost all targets considered in this paper were treated as single stars in DR3, which introduces an anomaly in their proper motion values. This limits the accuracy of our simulations. For systems with known orbits, as in our sample, this bias can be estimated and corrected. We follow the scheme below:

\begin{enumerate}
\item We simulate the system assuming zero proper motion over the DR3 time span.
\item We fit a SS model to these simulated data.
\item We subtract the fitted proper motion from the Gaia DR3 proper motion.
\end{enumerate}

The fitted proper motion corresponds to the apparent motion induced purely by the orbital motion. Subtracting this contribution from the Gaia DR3 proper motion removes the orbital bias and yields an estimate closer to the true barycentric proper motion of the system. We then apply this correction in our simulations to obtain the most realistic representation of the systems considered here.

\section{Validation \label{Validation}}

\subsection{BH3: comparison with Gaia DR4 epoch astrometry \label{Sec:BH3}}

To validate our forward simulations, we applied the pipeline to LS~II~+14~13, better known as Gaia BH3. This system is a good application as its orbit solution and astrometric data were published by \citet{Panuzzo_2024} before the official release of DR4.

To reproduce the BH3 astrometry, we directly adopt the parameters from the astrometry-only solution derived by \citet{Panuzzo_2024}, including the orbital elements, proper motion, and parallax. We used these parameters as input to our forward model to simulate the along-scan positions of BH3 at the observation epochs of the published data. We compared the simulated along-scan positions with the published BH3 measurements, which resulted in $\chi^2_{\rm red}$ of 0.65. This validates our binary model implementations. Unfortunately, we cannot perform the same validation for the VIM model, as no public Gaia data are available at the time of writing for a VIM object.

\subsection{AK For: comparison with Gaia NSS solution \label{Sec:AKFor}}

\begin{figure*}[hbt!]
  \centering \includegraphics[width=.8\hsize]{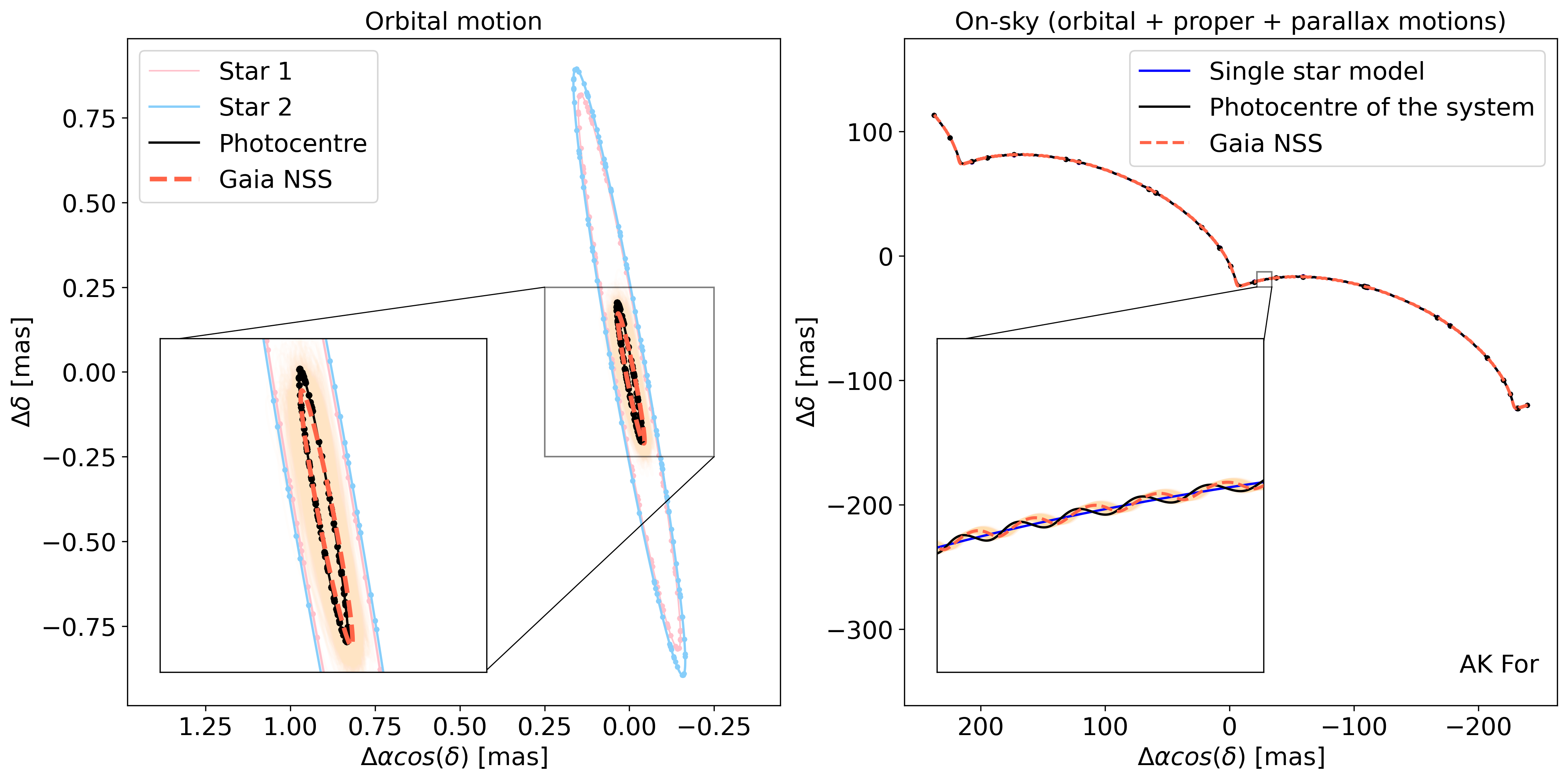}
  \caption{Photocentre motion for the AK~For binary system. Left panel: respective orbits of both components centred on the centre of mass and the resulting photocentric orbit. Right panel: on-sky motion illustrating the apparent trajectory including orbital, proper, and parallax motion. In both panels, the photocentre motion calculated using the literature values discussed in Sect.~\ref{Sec:AKFor} is shown as black full line. The simulation using the Gaia NSS solution is represented with an orange dashed line, with the shaded region indicating the uncertainty from 1000 Monte Carlo orbital realizations.}
  \label{AK_For}
\end{figure*}

The AK~For system, short-period double-lined spectroscopic binary, was one of the sources for which a full orbital NSS solution was derived in DR3. The system was part of a targeted search within the NSS processing. While most sources in the NSS pipeline were selected based on a poor goodness of fit (GoF), the targeted search included objects with previously known companions \citep{Arenou_2023}.

\citet{Gallenne_2023} combined interferometric astrometry with radial velocity measurements of both components, constraining orbital parameters, yielding model-independent distances as well as orbital parameters and component masses. This binary system provides an opportunity to assess the performance of Gaia astrometry when binarity is properly accounted for. For this, we simulate AK~For system by using orbital parameters from \citet{Gallenne_2023}.

To compute the photocentric semi-major axis $a_{\mathrm{ph}}$, we need the flux ratio in the Gaia $G$ band. As \citet{Gallenne_2023} report flux ratios in the $K$ band, we extrapolate these values to the $G$ band using Castelli-Kurucz model atmospheres \citep{Kurucz_2003} assuming solar metallicity for the stellar parameters listed in \citet{Gallenne_2023}. The resulting flux ratios and component magnitudes can be found in Table~\ref{tab:binary}.

In Table~\ref{tab:AK For}, we compare the parameters inferred from the NSS pipeline with those reported by \citet{Gallenne_2023}.
According to the latter paper, the system's orbit is nearly circular  and close to edge-on, making several parameters difficult to constrain from astrometry alone. In such configurations, degeneracies typically affect the longitude of the ascending node ($\Omega$, here shifted by 180$^\circ$), the argument of periastron ($\omega$, which differs between the two orbital solutions), and the periastron time $T_0$ (shifted by one third of the period).
The latter two parameters are completely degenerate in the case of a circular orbit, which explains the large uncertainties reported in \citet{Gallenne_2023}.
The main difference between the solutions lies in the eccentricity ($e$), which is close to zero in \citet{Gallenne_2023} but reaches 0.287 in the Gaia DR3 solution.
The Gaia DR3 NSS parallax agrees within uncertainties with the value reported by \citet{Gallenne_2023}, and the value from the Gaia main catalogue is likewise consistent within the error bars.
In terms of parallax accuracy, the difference between the binary and single-star models is only 0.01 mas.

\begin{figure}[hbt!]
  \centering \includegraphics[width=.98\hsize]{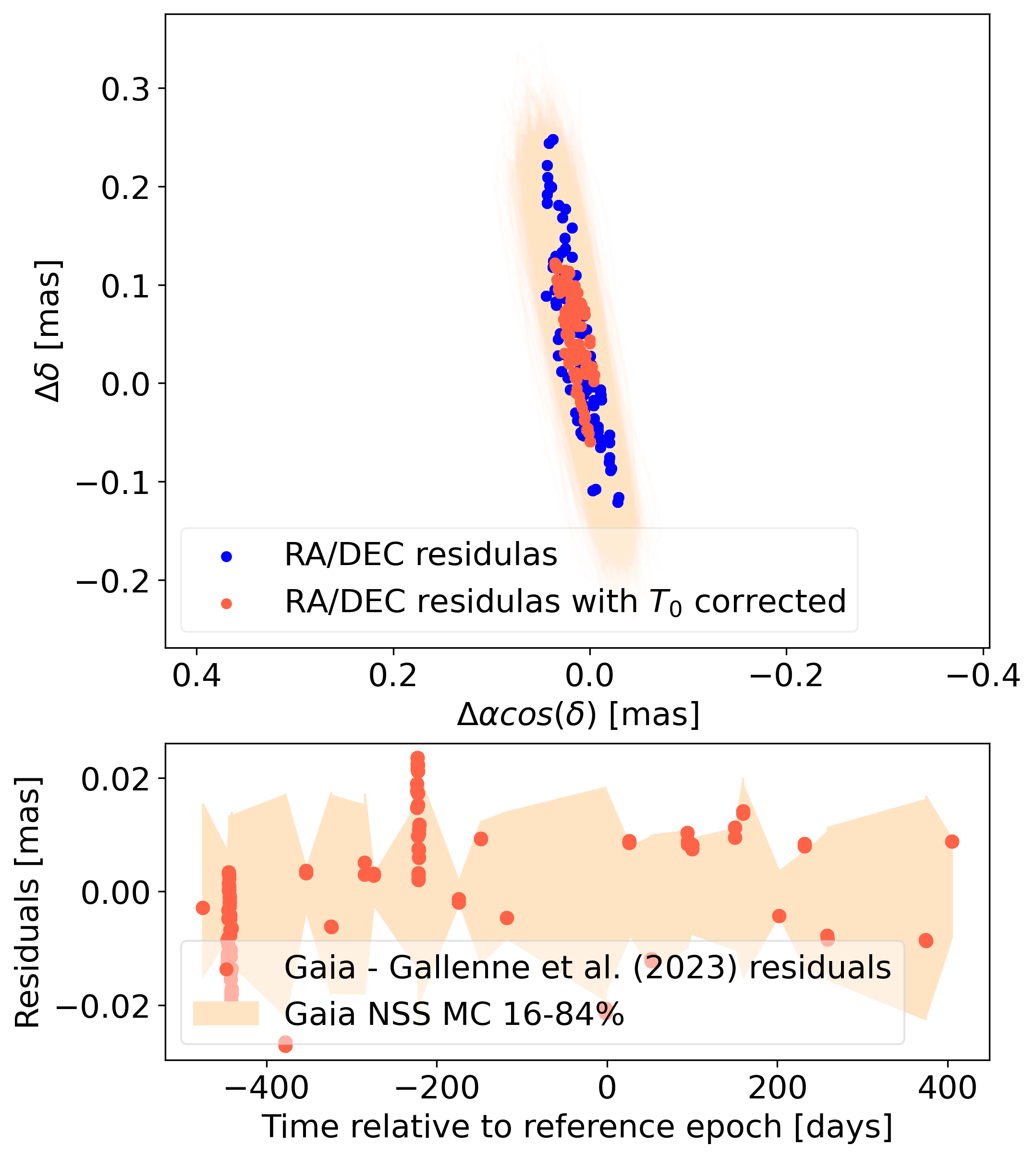}
  \caption{RA and DEC residuals for each point between the two parameter sets. The blue dots compare the original parameters from Table~\ref{tab:AK For}, while the orange ones account for the $T_0$ offset. The shaded region indicating the uncertainty from 1000 Monte Carlo orbital realizations of the full astrometric solution from Fig.~\ref{AK_For} is shown in pale orange. Bottom panel: $T_0$ offset-accounted AL residuals compared with confidence region of Gaia solution computed with its covariance matrix.}
  \label{AK_For_res}
\end{figure}

Despite the differences in the orbital parameters, the overall orbital geometry of the astrometric positions generated using the parameters from \citet{Gallenne_2023} remains very similar to that of the Gaia NSS solution (see left panel of Fig.~\ref{AK_For}). A closer inspection of the on-sky motion reveals a periodic offset between the trajectories, reflecting the differences discussed above: $T_0$ and $\omega$ are not well constrained in \citet{Gallenne_2023}'s near-circular orbit.
To further investigate the differences between the two orbital solutions, we plot the RA and DEC residuals between the Gaia solution and the solution based on the parameters from \citet{Gallenne_2023} (top panel of Fig.~\ref{AK_For_res}). However, the disagreement in $T_0$ between the models introduces a periodic shift, making a direct comparison difficult. To account for this, we shift $T_0$ for the solution from \citet{Gallenne_2023} and plot the corresponding residuals. After accounting for the difference in $T_0$, the residuals remain within the uncertainty region of the Gaia solution's orbit. We additionally verified that 88.4\% of the AL residuals between the Gaia and literature solutions fall within the 16th–-84th percentile range (bottom panel of Fig.~\ref{AK_For_res}).
We therefore conclude that the two solutions are statistically consistent, despite the differences in the individually fitted orbital parameters. This agreement supports both the Gaia NSS solution and our estimated photocentric orbit, while also illustrating the degeneracy between orbital solutions when the data do not fully constrain the orbit.

\subsection{Proper motion correction validation with BH3}

BH3 also provides a suitable test case for validating the proper-motion correction introduced in Sect.~\ref{correction}. The DR3 proper motion values and those reported by \citet{Panuzzo_2024} (DR4) are listed in Tab.~\ref{tab:pm_bh3}. The DR3 values represent the ''raw'' proper motions, affected by the orbital anomaly, whereas the DR4 values provide the corresponding ''true'' (centre of mass) proper motions, derived by accounting for binarity. The difference between the two sets reaches $-6.082 \pm 0.129$ mas\,a$^{-1}$ in right ascension and a negligible $-0.055 \pm 0.170$ mas\,a$^{-1}$ in declination. After applying our correction to the raw DR3 values, we obtain the corrected values listed in Table~\ref{tab:pm_bh3}, which show better agreement ($\sim 1.3 \sigma$) with those reported by \citet{Panuzzo_2024} with difference of $\Delta \mu_{\alpha^\star}=0.186 \pm 0.184$ and $\Delta \mu_{\delta}=-0.243 \pm 0.169$ mas\,a$^{-1}$. The remaining discrepancy is likely due to noise, which is not included in our correction procedure, as it is impossible to exactly reproduce the errors occurring in the real data, while bootstrapping would only affect the uncertainty of the correction, not its value.

This level of accuracy is nevertheless sufficient to produce more realistic simulations when only biased DR3 values are available. We therefore apply this proper motion correction within our pipeline to all targets that do not have unbiased proper motion measurements.

\begin{table}
\caption{Proper motion comparison for BH3.}
\label{tab:pm_bh3}
\centering
\setlength{\tabcolsep}{3pt}
\renewcommand{\arraystretch}{1.2}
\begin{tabular}{lccc}
\hline\hline
 & DR3 & DR4 & Corrected \\
\hline
$\mu_{\alpha^\star}$ [mas\,a$^{-1}$] & $-22.235^{\pm 0.062}$ & $-28.317^{\pm 0.067}$ & $-28.503^{\pm 0.122}$ \\
$\mu_{\delta}$   [mas\,a$^{-1}$]    & $-155.276^{\pm 0.059}$ & $-155.221^{\pm 0.111}$ & $-155.033^{\pm 0.110}$ \\
\hline
\end{tabular}
\end{table}

\section{Synthetic data \label{Sec:sample}}

In this section, we present six binary Cepheids and additional 11 non-variable binaries and discuss their simulations. We adopt the proper motion values from DR3 and correct them using the method described in Section \ref{correction}. Because the Gaia G band has a similar mean wavelength to the V band, we use this approximation in the simulations.

\begin{figure*}[hbt!]
  \centering \includegraphics[width=.99\hsize]{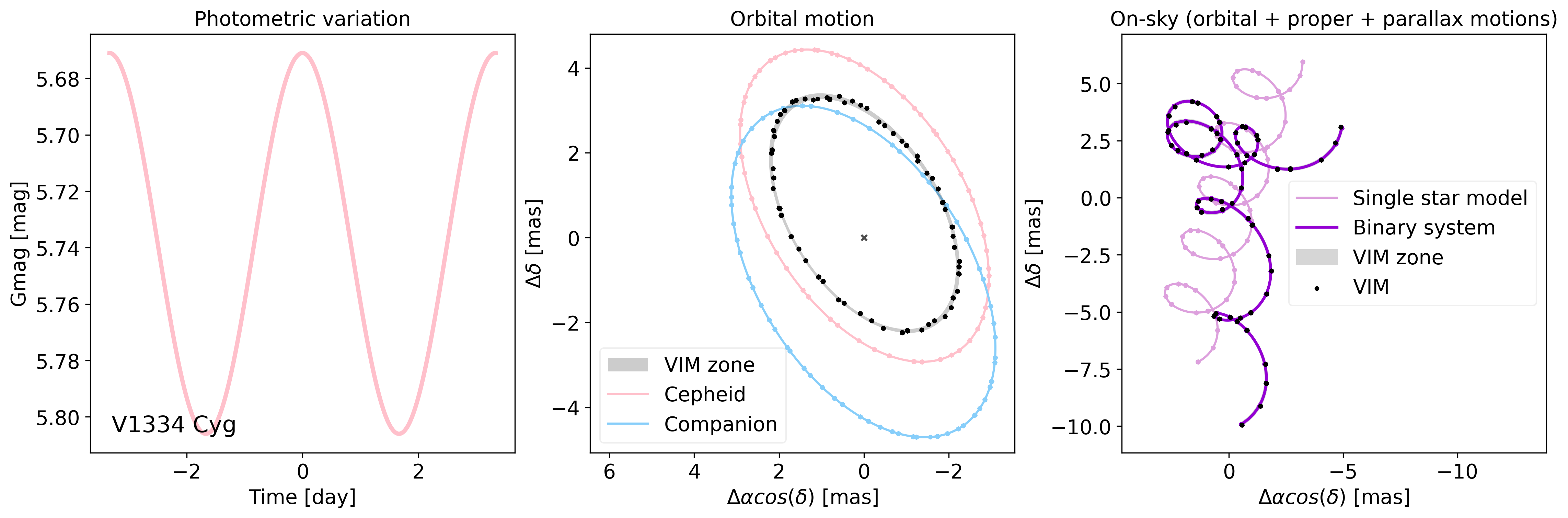}
  \caption{Left panel: reconstructed photometric variability of the Cepheid V1334 Cyg  in G band from Gaia DR3 Cepheid catalogue \citep{DR3var}. Middle: component orbits around the centre of mass (grey cross) and the corresponding photocentric orbit, together with the grey zone indicating the VIM amplitude (see Fig.~\ref{AWPer} for a clearer example). Right panel: single-star sky path with the same proper motion and parallactic motion as V1334 Cyg but without binarity (light purple). The binary-system curve includes orbital motion but no pulsation (dark purple). Black dots represent the full system, including pulsation. The black dots visually lie on the binary curve because the VIM effect remains negligible in this case.}
  \label{V1334Cyg} 
\end{figure*}

\paragraph{V1334 Cyg.} V1334 Cyg is a first-overtone Cepheid with a B7V companion \citep{Evans_1995}. \citet{Gallenne_2018} resolved the system spectroscopically and interferometrically and achieved a distance precision of one percent ($d = 720.35 \pm 7.84$ pc, corresponding to $\varpi = 1.388 \pm 0.015$ mas). We adopt their orbital parameters and parallax for our simulations.  \citet{Evans_1995} estimated a magnitude difference of $\Delta V = 2.18$ between the two components. Given a system magnitude of 5.72~mag in DR3, this corresponds to a companion magnitude of 8.04~mag in G band. Figure~\ref{V1334Cyg} shows the simulated photocentric motion of the system over the DR4 time span.

\paragraph{SU Cyg.} SU Cyg is a triple system composed of a Cepheid and companions Ba (B7.5V) and Bb (later than A0V) \citep{Gallenne_2025}. \citet{Gallenne_2025} performed an analysis similar to that of \citet{Gallenne_2018} and obtained the most precise Cepheid distance to date, with a precision of 0.5\% ($d = 926.3 \pm 5.0$ pc, $\varpi = 1.080 \pm 0.006$ mas). In their fit, they treated subsystem B as a single source, assuming its photocentric wobble in the H band to be negligible.
We performed a similar calculation in the V band using their parameters and their Equation~6 to estimate the contribution of subsystem B. We found that the wobble remains below 12 $\mu$as, comparable to the per-transit uncertainty. We also carried out an approximate simulation including all three stars, which resulted in additional perturbations smaller than 0.75 $\mu$as in $w_{\mathrm{AL}}$. We therefore conclude that the B companion subsystem can be treated as a single object.

According to \citet{Gallenne_2025}, the magnitude difference in V between the B subsystem and the Cepheid is about 3 mag. With a Gaia magnitude of 6.80 mag for the system, this gives a companion magnitude of 9.86 mag. The corresponding DR4 simulation is presented in Fig.~\ref{SUCyg}.

\paragraph{V350 Sgr.} This system was also studied using interferometry combined with radial velocities \citep{Evans_2025}. However, no ultraviolet spectra were available to isolate the companion lines, preventing a full determination of system parameters and leaving a degeneracy between masses and distance. The authors estimated a distance of $d = 986 \pm 48$ pc ($\varpi = 1.014 \pm 0.049$ mas) based on assumed stellar masses. The inverse approach, assuming the Gaia DR3 distance, led to unexpectedly large masses, suggesting that the Gaia distance may be overestimated.
\citet{Evans_2025} reported a companion magnitude of 11.13. Instead, to avoid inconsistencies between photometric zero-points, we adopt a Cepheid--companion magnitude difference of 3.62, which yields a companion magnitude of 10.90 using the Gaia system magnitude of 7.25. The visualisation of the DR4 simulation can be found in Fig.~\ref{V350Sgr}.

\paragraph{U Aql and S Mus.} \citet{Gallenne_2019} conducted a multiplicity survey of 16 Galactic Cepheids. We retained only two systems for which preliminary orbits were produced. As in the case of V350 Sgr, no companion spectral lines were detected, preventing independent distance determinations. Instead, the authors adopted distances from the period–luminosity relation \citep{Storm_2011}:
U Aql: $592 \pm 19$ pc ($\varpi = 1.689 \pm 0.054$ mas),
S Mus: $858 \pm 17$ pc ($\varpi = 1.166 \pm 0.023$ mas).
To estimate companion magnitudes, we assume spectral types from \citet{Gallenne_2019} (A0V for U Aql B and B6V for S Mus B) and use Table 5 of \citet{Mamajek_2013} to derive V-band magnitudes at the assumed distances: 9.85 and 9.12 mag, respectively. Fig.~\ref{UAql} and Fig.~\ref{SMus} present these simulations.

\begin{figure*}[hbt!]
  \centering \includegraphics[width=.99\hsize]{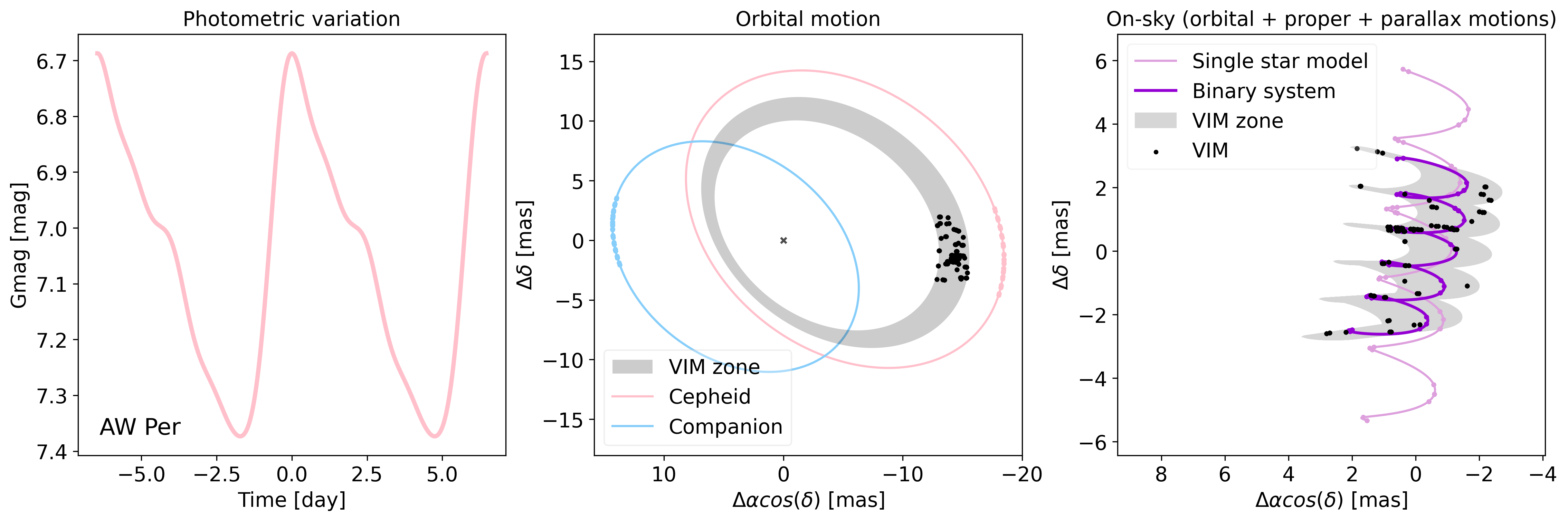}
  \caption{Figure of AW Per similar to V1334 Cyg (Fig.~\ref{V1334Cyg}). In this case, the black dots show a larger amplitude, which corresponds to the VIM effect.}
  \label{AWPer} 
\end{figure*}

\paragraph{AW Per.} AW Per is a triple system \citep[or possibly quadruple, with an additional companion resolved by Gaia;][]{Kervella_2019b} with a main orbital period of 38 years. \citet{Evans_2024} studied this system but did not derive an independent parallax. Nevertheless, we include it in our sample because it provides a clear example of a strong VIM effect. We therefore adopted the parallax of $1.092 \pm 0.029$ mas from the Gaia DR3 catalogue with the \citet{Lindegren_2021} correction applied for the modelling.

Similarly to the case of SU Cyg, the Cepheid's companion forms a subsystem composed of Ba and Bb. \citet{Evans_2024} separated their spectral lines in the ultraviolet domain. They found two nearly identical B7V stars with a flux ratio $f(Ba)/f(Bb) = 1.05$. We lack sufficiently precise orbital parameters to estimate the photocentric wobble of this subsystem, but their similar properties suggest that $a_{\mathrm{ph,B}}$  will be even smaller than in SU Cyg. We therefore treat this subsystem as a single object, as in the referenced study.
From Table 2 of \citet{Evans_2024}, the B subsystem is 2.09 mag fainter than the Cepheid in V band. Given a Gaia magnitude of G=7.09 for the system, this corresponds to a companion magnitude of G=9.33. Fig.~\ref{AWPer} shows the DR4 simulation of AW~Per.

\paragraph{Non-pulsating binaries.} To our Cepheid sample, we add a sample of 11 systems containing photometrically stable stars. These systems were studied and treated in the same manner as AK~For, described in Section~\ref{Sec:AKFor}, and are taken from \citet{Gallenne_2016} and \citet{Gallenne_2023}. Unlike AK~For, they were treated as single stars in DR3.

\section{Impact of binary motion and VIM effects on Gaia astrometric parameters in our sample \label{Sec:analysis}}

\subsection{Proper motion \label{Sec:propermotion}}

\begin{figure*}[hbt!]
  \centering \includegraphics[width=.99\hsize]{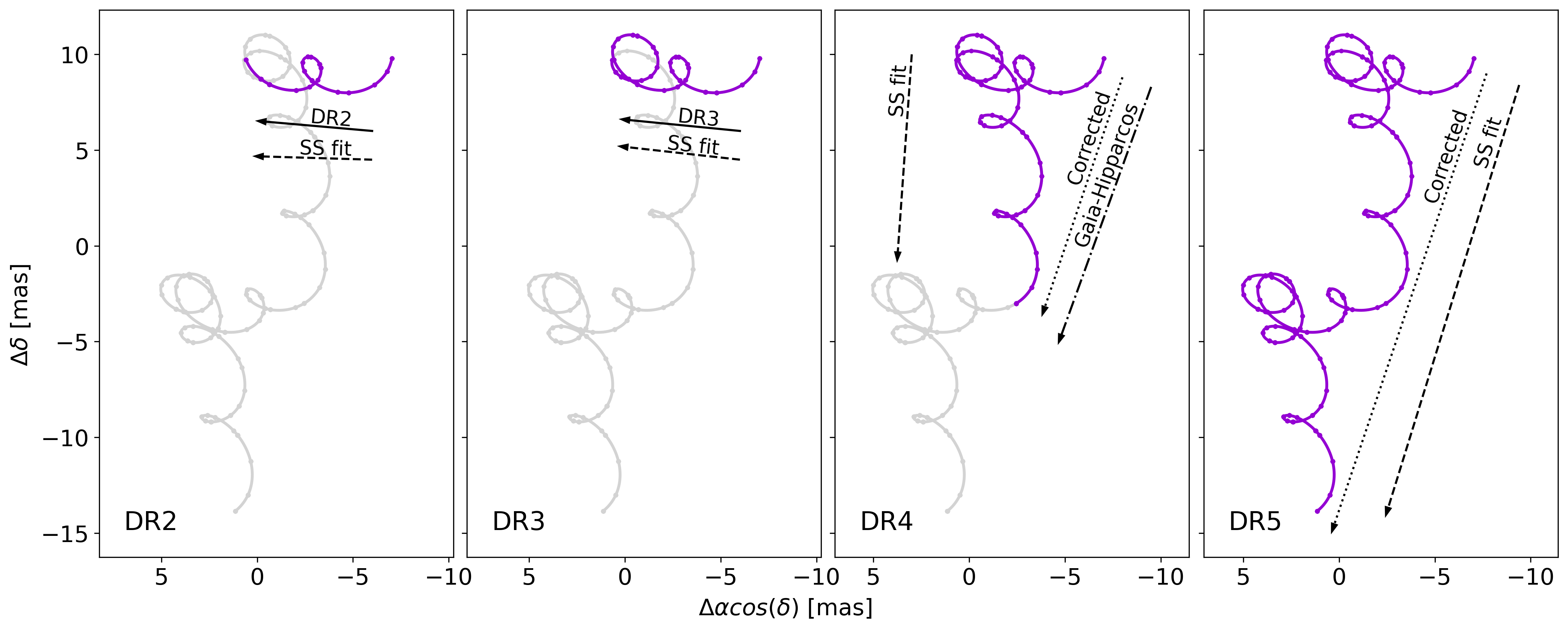}
  \caption{Simulated photocentric motion of V1334 Cyg over DR2, DR3, DR4, and DR5. The colored segments correspond to the positions covered in each respective release, while the gray curve shows the predicted full DR5 trajectory. Arrows indicate the direction of proper motion, scaled to the motion measured in each release. Solid arrows represent the proper motion reported by Gaia, while dashed arrows correspond to the values obtained from SS fits to our simulated data. The dotted line shows the centre-of-mass motion corrected using the method from Section \ref{correction}, and the dash-dotted line corresponds to the Gaia-Hipparcos proper motion from \citet{Kervella_2022}.
  }
  \label{PM_V1334 Cyg}
\end{figure*}

As previously mentioned, all our Cepheids were treated as single stars in DR3, which impacted their proper motion estimates. Here, we investigate this impact in our sample. We first focus on V1334 Cyg and then on the other Cepheids and non-pulsating binaries.

In Fig.~\ref{V1334Cyg}, we saw that the forward simulation for a binary system diverges significantly from the SS simulation computed using the same parallax and proper motion parameters. Fig.~\ref{PM_V1334 Cyg} illustrates how the motion of V1334 Cyg appears in different Gaia releases. The orbital period (approximately 63 months) of this system is comparable to the DR4 duration (66 months), so previous releases observed only a fraction of the orbit. This leads to proper motion values strongly affected by orbital contamination. As described in Section \ref{correction}, our simulations use a PM correction that, in this case, gives a proper motion vector of $\mu_{\alpha^\star}=0.78 \pm 0.11$ and $\mu_{\delta}=-2.31 \pm 0.13$ mas a$^{-1}$. These values are comparable with the Gaia--Hipparcos proper motion from \citet{Kervella_2022} of $\mu_{\alpha^\star}=0.90 \pm 0.01$ and $\mu_{\delta}=-2.50 \pm 0.01$ mas a$^{-1}$.

We also performed an SS fit on the simulated data corresponding to the Gaia data releases DR2 to DR5. As can be seen in Fig.~\ref{PM_V1334 Cyg}, the fitted vectors reproduce those published in DR2 ($\mu_{\alpha^\star}=5.55 \pm 0.10$, $\mu_{\delta}=0.33 \pm 0.11$ mas a$^{-1}$) and DR3 ($\mu_{\alpha^\star}=2.29 \pm 0.06$, $\mu_{\delta}=0.23 \pm 0.08$ mas a$^{-1}$). Based on this approach, we predict a proper motion of $\mu_{\alpha^\star}=0.15 \pm 0.33$ and $\mu_{\delta}=-1.97 \pm 0.37$ mas a$^{-1}$ for DR4 if the source were to be treated as a single star. It is also noticeable that, for DR5, with a duration of almost 126 months, corresponding to twice the orbital period of the system, the proper motion derived from the SS fit is no longer affected by the PM anomaly.

For the other Cepheids, the situation is similar, although less significant than for V1334~Cyg. A visual comparison between the Gaia DR3, Gaia--Hipparcos, SS fit, and corrected PM vectors is shown in Fig.~\ref{PMs_cepheids}, while the corrected values are listed in Tab.~\ref{tab:pm_corrections}.

\begin{figure}[hbt!]
  \centering \includegraphics[width=.99\hsize]{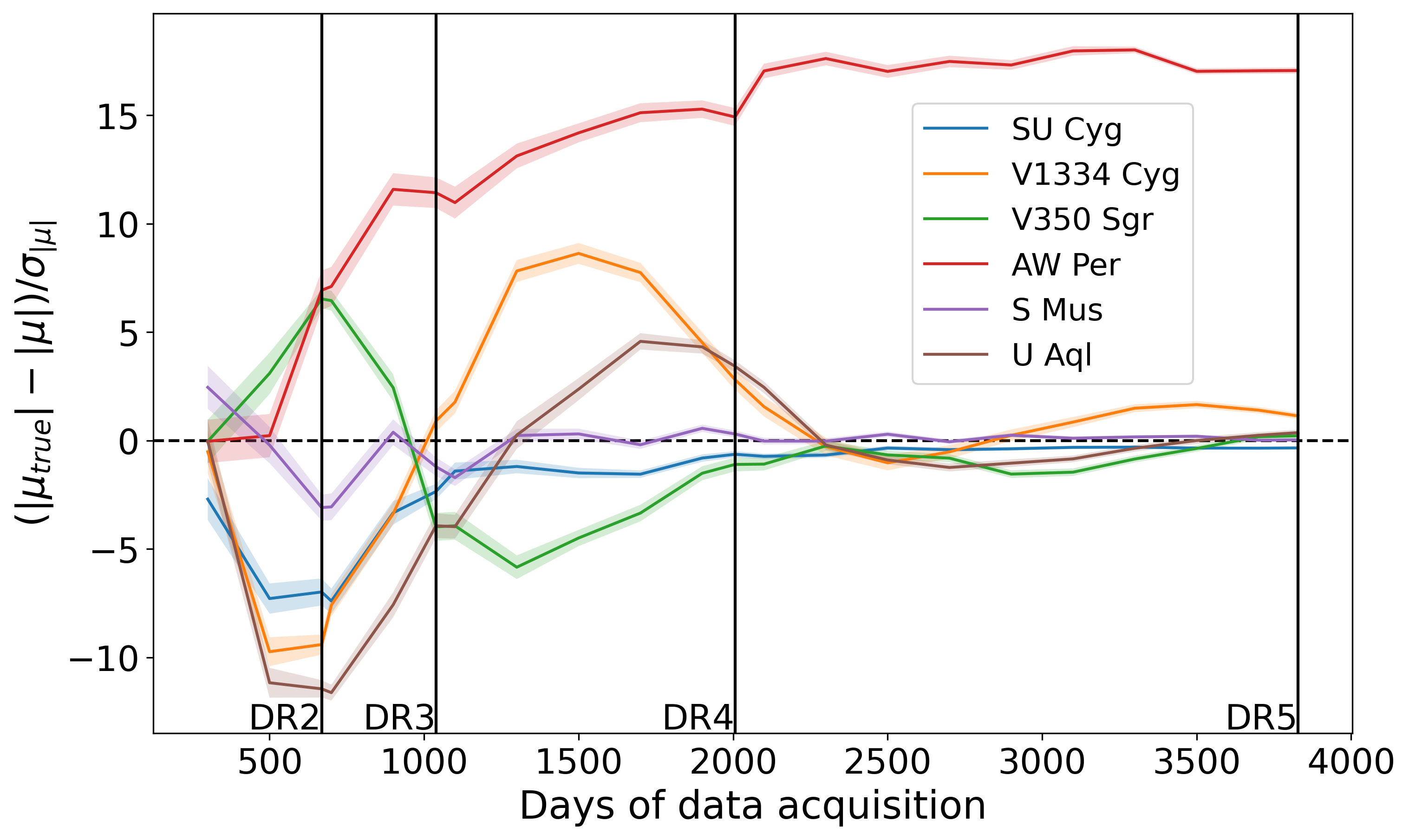}
  \caption{Difference between the expected PM values (input values of the simulation) and the fitted values for the Cepheid sample. The days on the $x$-axis correspond to the baseline used for the SS fit.}
  \label{pms_evolution_cepheids}
\end{figure}

We can further study how the PM anomaly varies with time. For this, we sampled the Gaia time span starting at 300 days from the start of the scientific mission and continuing to the end of the mission, with a step of 200 days and forced stops at Gaia data releases. We then considered data samples including all data from the start of the scientific mission up to the considered date and fit an SS model to each sample, with a bootstrap of 1000 realisations. We compared the modulus of the fitted proper motion to the input value of the simulation in terms of $\sigma$. The resulting plot can be seen in Fig.~\ref{pms_evolution_cepheids}. We reached the same conclusion for V1334~Cyg as previously: the accuracy improves with time. The same trend is observed for the other Cepheids, except AW~Per. The behaviour of AW Per can be explained by its long orbital period: with time, Gaia covers an increasing fraction of the orbit, which distorts the perceived trajectory.

\begin{figure}[hbt!]
  \centering \includegraphics[width=.99\hsize]{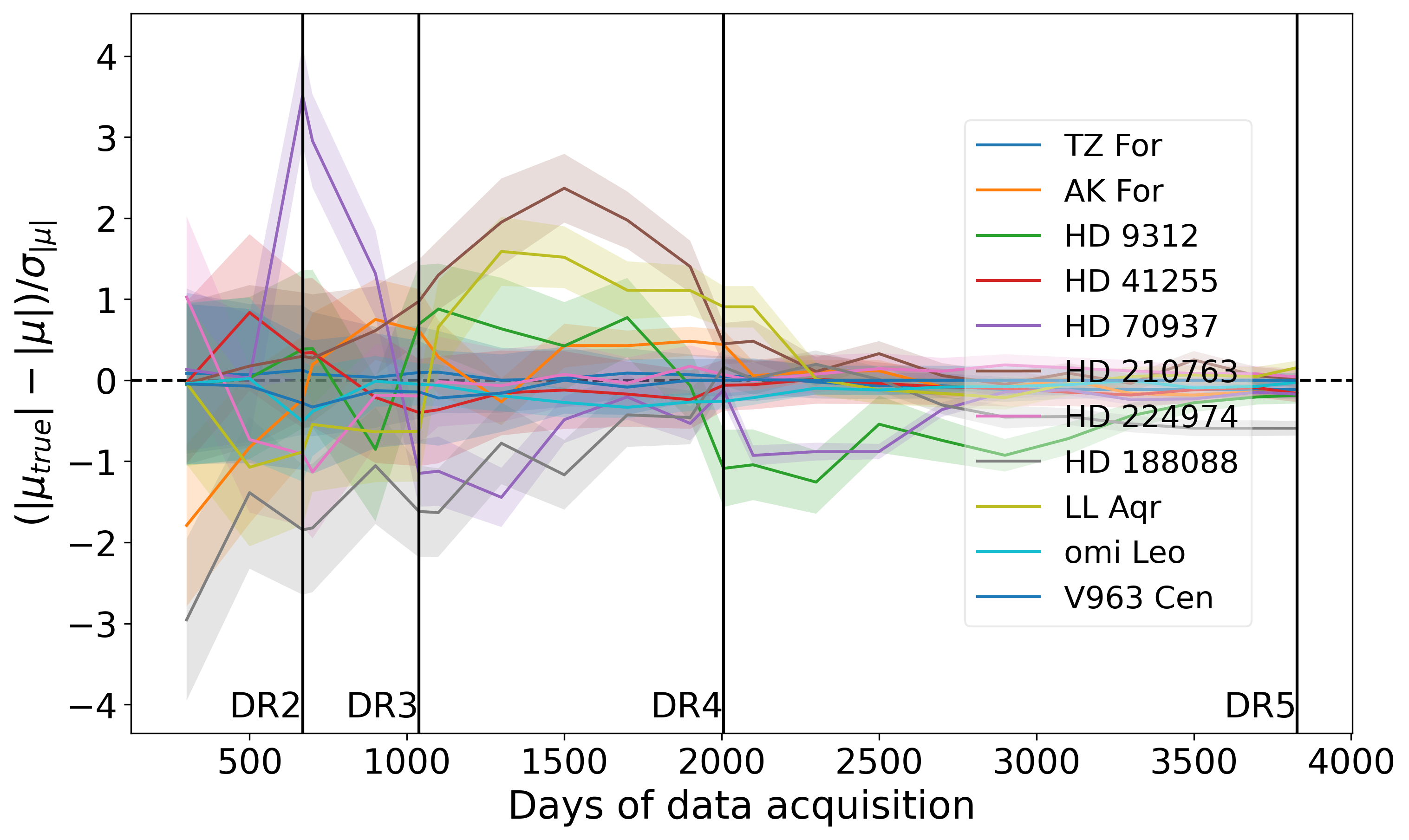}
  \caption{Difference between the expected PM values (input values of the simulation) and the fitted values for the non-pulsating binary sample. The days on the $x$-axis correspond to the baseline used for the SS fit.}
  \label{pms_evolution_binaries}
\end{figure}

For the non-pulsating binaries, we performed the same analysis, and the result can be seen in Fig.~\ref{pms_evolution_binaries}. We observed the same trend, with the error decreasing with time. From this, we conclude that the proper motion value becomes more accurate with time for systems with orbital periods shorter than the Gaia time coverage, even when using an SS model.

\subsection{Parallax \label{Sec:parallax}}

In this section, we are interested in the possible perturbations introduced in parallax measurements by the VIM effect or binary motion. Starting with the VIM effect, for five out of six Cepheids, the VIM amplitude is negligible: the $d_{\rm var}$ parameter does not exceed 0.20 mas, which is comparable to the per-CCD uncertainty. The one exception is AW Per, which has a VIM amplitude of 2.23 mas, about twice the expected parallax value, due to a favourable combination of stellar magnitudes and a large separation between the components. The second particularity of this star is its long period: the DR4 time span covers only about 15\% of the orbit. This small portion of the orbit can be approximated by a linear displacement, which is easily absorbed by the proper motion anomaly, isolating the influence of the VIM effect on the parallax.

To investigate the error introduced by the VIM effect in the parallax measurement of AW Per, we first fitted these simulated data with SS, VIMF, VIML, and VIMA models (see App.~\ref{annex:vim} and App.~\ref{annex:fitting}) and obtained reduced $\chi^2_{\rm red}$ values of 3.70, 1.29, 1.13, and 1.11, respectively, using a bootstrap of 10 000 realisations. The retrieved proper motion values from the VIM model fits agree with those from the SS fit. The VIMA model does not significantly improve the VIML model, as expected, because the observed motion is well approximated by a linear trend given the limited orbital coverage. The VIML model provides both the SS parameters and the VIM-specific ones (see equations in App.~\ref{annex:vim}); however, the centre-of-mass position and proper motion can only be estimated with an acceleration model or higher-order model \citep{Wielen_1996}. We performed a similar fitting process for DR5 for this target. Unfortunately, the DR5 time coverage was still insufficient to justify an accelerated VIM model, even though it corresponds to approximately 27\% of the orbit. The $\chi^2_{\rm red}$ improved from 2.42 to 2.40 between the VIML and VIMA models for DR5.

We compared the parallax derived from the SS model with those from the VIML and VIMA models for the DR4 simulation fit. Because no independent parallax measurement exists for this system, we used the Gaia DR3 value as a reference (see Section \ref{Sec:sample}), although it may not be accurate in absolute terms. Since we simulated the system using this parallax, it represents the “true” value in our simulations. The SS model gave 0.88 $\pm$ 0.04 mas, while the VIML and VIMA models gave 1.11 $\pm$ 0.02 mas, in agreement with the input value of 1.092 $\pm$ 0.029 mas. This case shows that the VIM effect can introduce significant scatter and strongly affect the resulting parallax.

\begin{figure}[hbt!]
  \centering \includegraphics[width=.99\hsize]{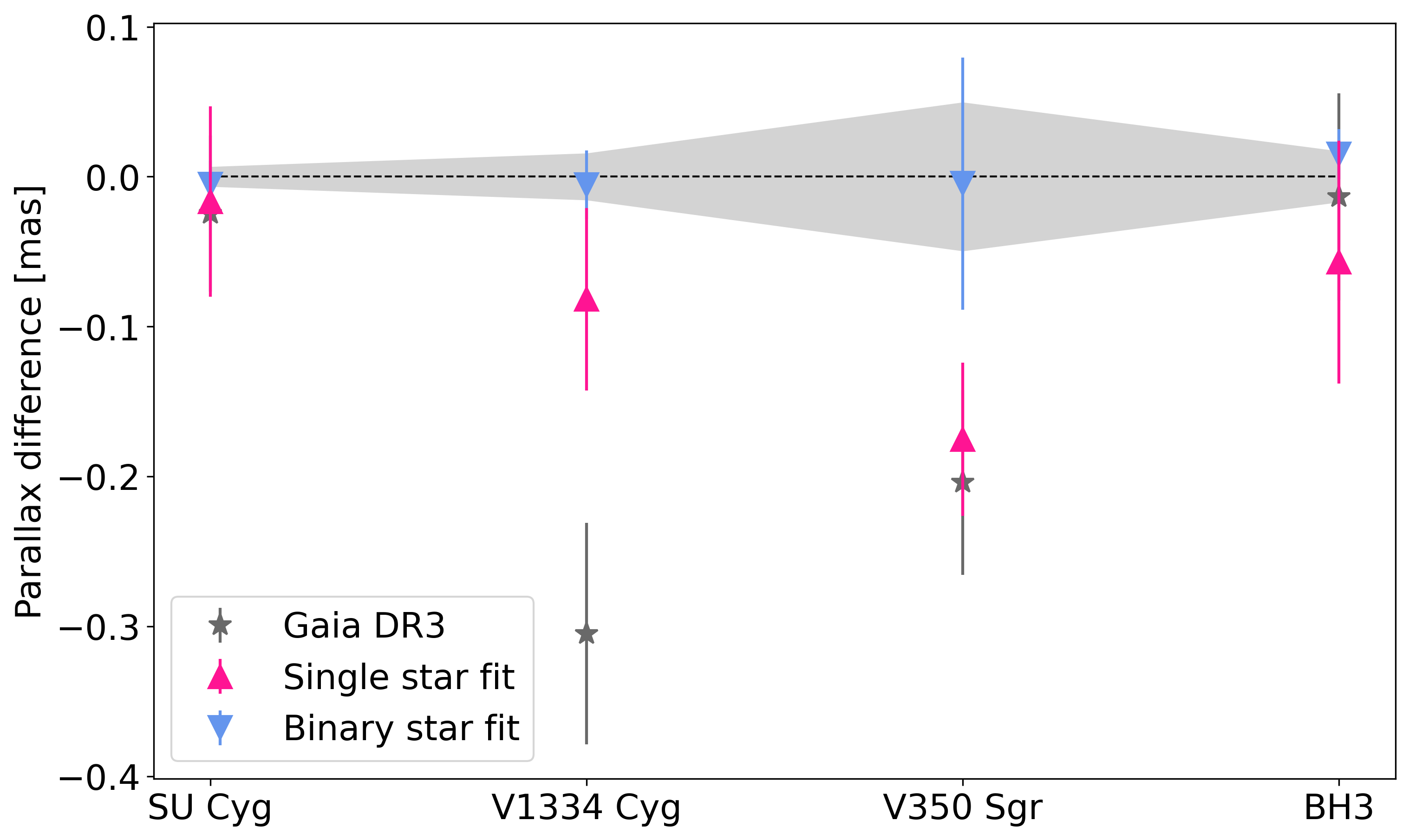}
  \caption{Parallax comparison for three Cepheid targets and BH3. The zero line represents the reference value from Table~\ref{tab:plx_cepheid}. The y-axis shows the difference between the Gaia or fitted value and the reference (Gaia/fit - ref). The grey shaded region corresponds to the $1\sigma$ uncertainty on the reference value. }
  \label{plx_cepheids}
\end{figure}

In the case of the remaining five Cepheids, the parallax is still affected by binary motion. We first want to determine whether the Gaia DR3 parallax can be explained by the presence of the companion. For this, we consider three Cepheids with precise distance measurements to compare our retrieved values from the simulations with Gaia DR3. V1334 Cyg and SU Cyg have fully model-independent distances, while V350 Sgr has a distance derived from an assumed mass, which is reflected in its uncertainty. The other three Cepheids in our sample do not have sufficiently precise independent distance determinations and therefore cannot be used for this purpose. To increase the sample size, we also include BH3 to compare the DR3 single-star solution with the DR4 analysis that accounts for binarity.

We simulate the DR3 along-scan observations of these four targets over 1000 bootstrap realisations and fit SS and BS models to them. The BS model fitting is performed with the \texttt{kepmodel} Python package \citep{kepmodel}. The SS fitting should be able to reproduce the DR3 values, as it simulates the five-parameter model treatment in Gaia. Figure~\ref{plx_cepheids} shows the parallax difference between the Gaia DR3 values and the reference values (grey star symbols). Single star fits generally agree with the DR3 values, except for V1334 Cyg, which likely suffers from additional sources of contamination. We also note that the BS model shows a systematic improvement over the SS model.

\begin{figure*}[hbt!]
  \centering \includegraphics[width=.99\hsize]{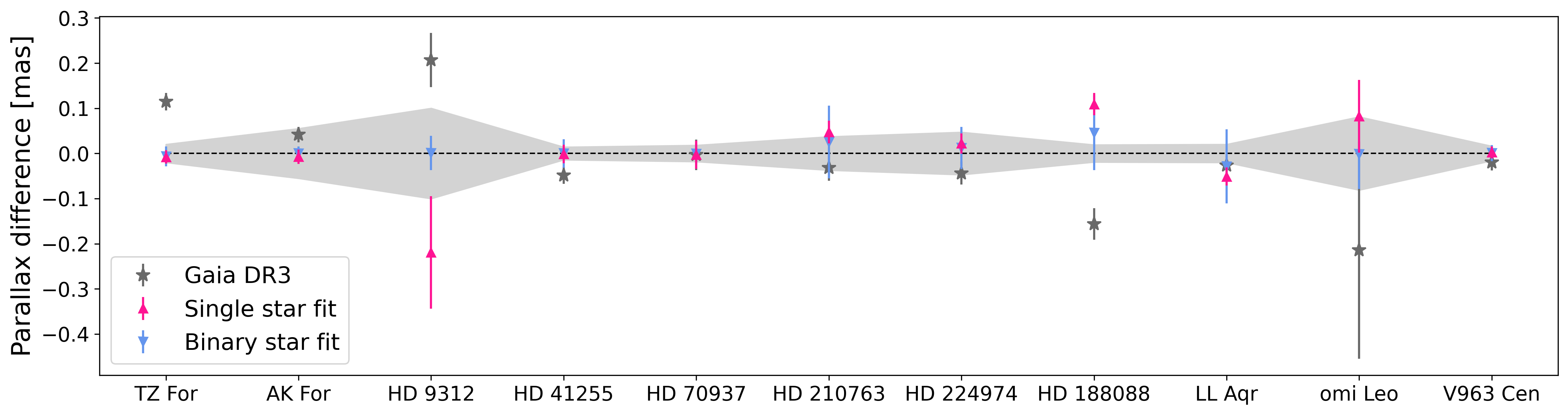}
  \caption{Parallax comparison for the non-pulsating binary sample. The zero line represents the reference value from Table~\ref{tab:plx_binary}. The y-axis shows the difference between the Gaia or fitted value and the reference (Gaia/fit - ref). The grey shaded region corresponds to the $1\sigma$ uncertainty on the reference value.}
  \label{plx_binaries}
\end{figure*}

We performed a similar analysis for the sample of 11 simple binaries, as all of them have independent parallax measurements. In Fig.~\ref{plx_binaries}, we do not observe the same behaviour as in our sample of three Cepheids and BH3: we do not find the same correspondence between the Gaia DR3 values and the SS fits. This results from the small contribution of binary motion to the astrometric signal, due to the small photocentric orbits, which act at a level comparable to the astrometric noise. As a result, the error in the DR3 parallax is not dominated by the presence of a companion. The parallaxes derived from the BS model are generally equal to or more accurate than those obtained with the SS model, but the difference remains small. This is also supported by the agreement between the parameters of AK For derived from the Gaia NSS solution and those from the Gaia main catalogue. We therefore conclude that, for such systems with small photocentric orbits, binary motion does not introduce a significant bias in the parallax.

\begin{figure}[hbt!]
  \centering \includegraphics[width=.99\hsize]{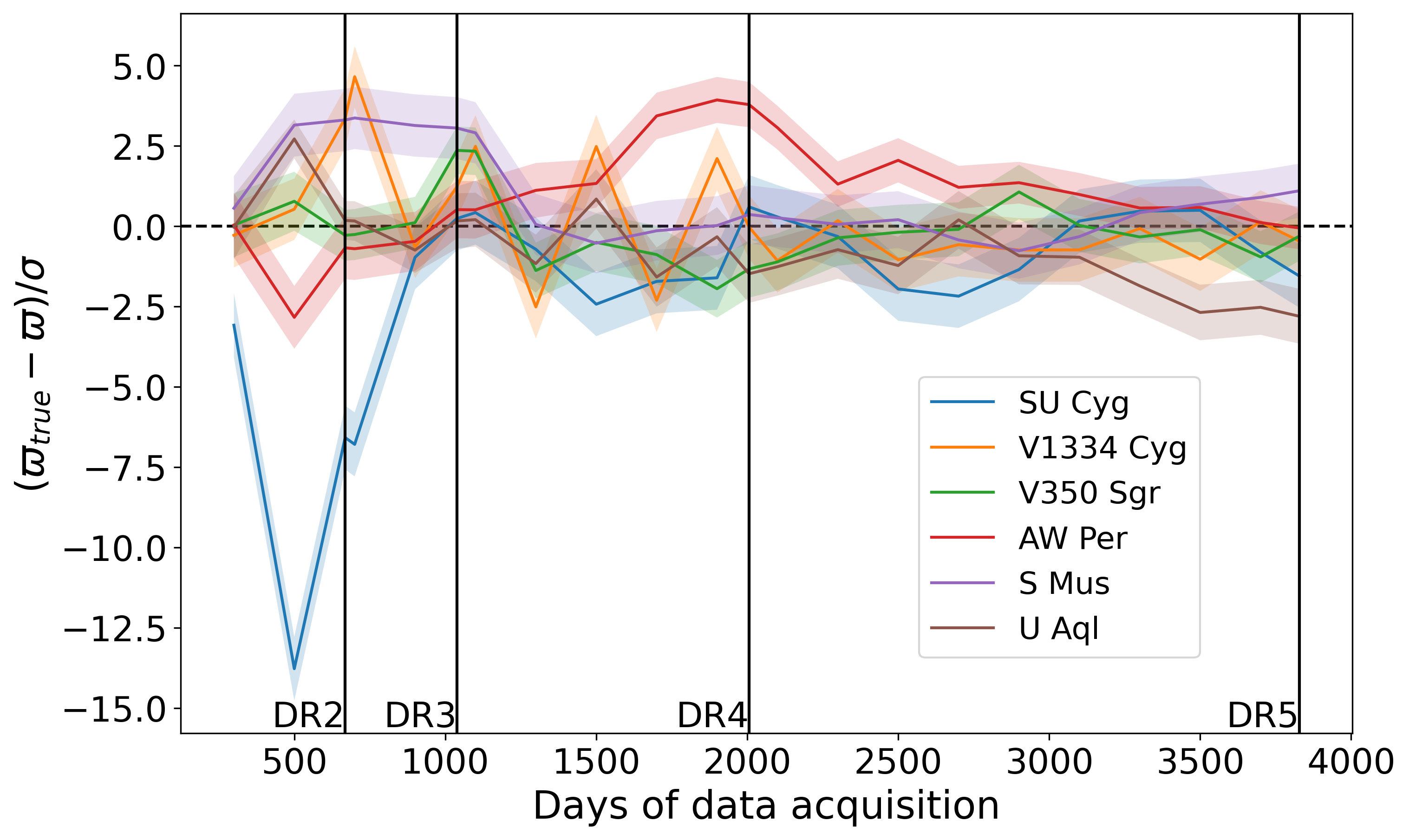}
  \caption{The difference of parallax values between the expected (the input of the simulation) and the fitted for Cepheid sample. Days in x-axis correspond to the baseline used for the SS fit.}
  \label{plxs_evolution_cepheids}
\end{figure}

We then perform an analysis following the prescription in Section~\ref{Sec:propermotion}, but applying it to parallax measurements for the six Cepheids. In Fig.~\ref{plxs_evolution_cepheids}, we do not observe a particular global bias that would systematically underestimate or overestimate the parallax value. Rather, we observe general scatter that does not decrease with time. The larger time coverage does not cancel the contribution of the parallax. The scatter is 3.63, 1.20, 1.90, and 1.34$\sigma$ from DR2 to DR5, respectively, or, in terms of percentage of the true value, 21.7\%, 8.9\%, 12.1\%, and 7.2\%. This dispersion in parallax translates into magnitude perturbations of 3.33, 1.71, 2.32, and 1.84$\sigma$, respectively. Considering the BS model for DR4 and DR5 brings these values down to 0.36$\sigma$ for both, or to 2.10\% and 0.55\% for the parallax, respectively. This reduces magnitude error to 0.67$\sigma$ for both DR4 and DR5.

\begin{figure}[hbt!]
  \centering \includegraphics[width=.99\hsize]{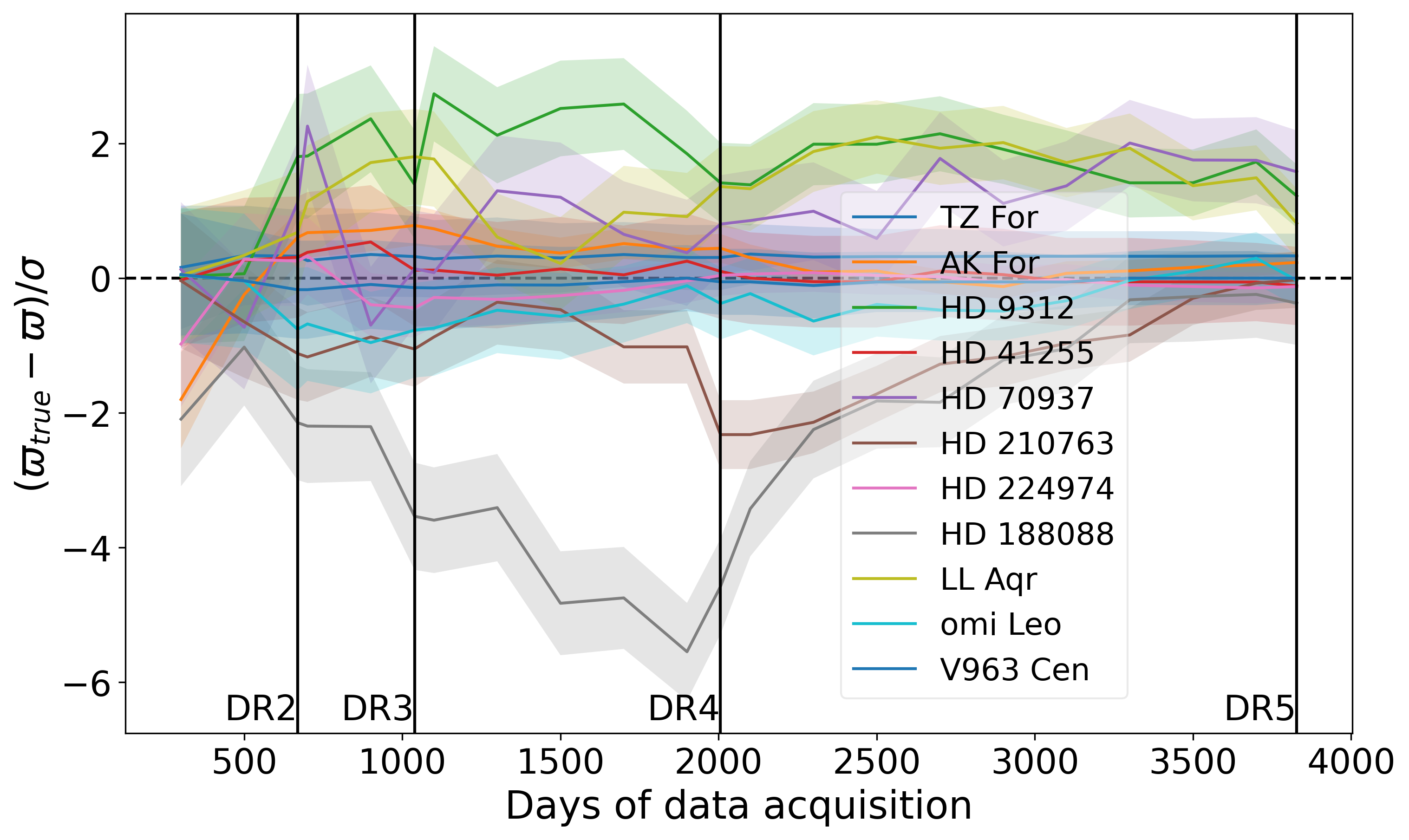}
  \caption{The difference of parallax values between the expected (the input of the simulation) and the fitted for non-pulsating binary sample. Days in x-axis correspond to the baseline used for the SS fit.}
  \label{plxs_evolution_binaries}
\end{figure}

In the case of non-pulsating binary systems (see Fig.~\ref{plxs_evolution_binaries}), the scatter is 1.09, 1.42, 1.75, and 0.62$\sigma$ from DR2 to DR5, respectively, remaining relatively stable and then sharply decreasing for DR5. Including the binary model for DR4 and DR5 reduces it to 0.08$\sigma$ for both.

Looking at each star individually in Figs.~\ref{plxs_evolution_cepheids} and \ref{plxs_evolution_binaries}, we see that the error varies significantly with time, oscillating in an unpredictable way. For example, V1334~Cyg and U~Aql, despite having comparable periods, do not show correlated parallax error evolution. The final parallax value depends on many parameters of the system and their relationships with each other. In order to achieve the best parallax precision for binary Cepheids, it is therefore more suitable to consider each case individually.

\section{Conclusions \label{Conclusion}}

The upcoming fourth \textit{Gaia} data release will be particularly important for binary detection thanks to the public release of epoch astrometry. In this article, we presented \texttt{vimbinary}, a Python tool designed to model Gaia DR4-like epoch astrometry. It can be used both before the DR4 release to generate realistic simulations and after the release to fit and interpret the observed data.

We showed that, for the considered systems, VIM and binarity effects affect parallax measurements. The accuracy does not necessarily improve with successive releases, but considering the appropriate model can significantly improve the accuracy of distance determinations. Accounting for these effects with an appropriate model is therefore expected to improve the calibration of the period--luminosity relation with Gaia DR4, both directly and through a better characterisation of the projection factor used in the parallax-of-pulsation method for Cepheid distance determinations \citep{Nardetto_2023}. Together, these improvements will contribute to a more accurate calibration of the local H$_0$ measurement.

We note that variability-induced motions are likely to become more important in the case of asymptotic giant branch (AGB) stars due to their large photometric variability. In addition, their surface convection introduces an extra contribution to the photocentric motion. Convection-induced variations were shown to be non-negligible by \citet{Beguin_2024}. Combined with the large photometric variability amplitudes of these stars, convection is therefore expected to strongly affect the astrometric solution. We will address this question in a second paper \citep{Decin_2026}.

In addition, these photocentre-limited observations could be complemented with ground-based observations to resolve the system, measure the component separation, and break the flux--mass degeneracy, allowing the determination of the component masses. This could help resolve the discrepancies between values derived from evolutionary models and those inferred from pulsation calculations for Cepheid masses \citep[current attempts to resolve this issue by including stellar rotation are still debated,][]{2026arXiv260514128S}, as well as improve mass determinations in binary systems in general.

\begin{acknowledgements}
      This work received funding from ANR Unlock-pfactor (ANR-23-CE31-0009). In addition, this research is supported by the European Research Council (ERC) under the European Union’s Horizon 2020 research and innovation program (grant agreement No. 951549 - UniverScale). PK acknowledges support from the Polish-French Marie Skłodowska-Curie and Pierre Curie Science Prize awarded by the Foundation for Polish Science.
      LD acknowledges support from the FWO grant G0B3823N,
      the FWO grant G099720N, the KU Leuven C1 excellence grant MAESTRO
      C16/17/007, and the KU Leuven Methusalem SOUL grant METH/24/012.
      AG acknowledges support from the Agencia Nacional de Investigaci\'on y Desarrollo (ANID) through FONDECYT Regular grant 1241073.
      LB acknowledges support through the European Space Agency (ESA) Research Fellowship in Space Science.
      This work has made use of data from the European Space Agency (ESA) mission {\it Gaia} (\url{https://www.cosmos.esa.int/gaia}), processed by the {\it Gaia} Data Processing and Analysis Consortium (DPAC,\url{https://www.cosmos.esa.int/web/gaia/dpac/consortium}). Funding for the DPAC has been provided by national institutions, in particular the institutions participating in the {\it Gaia} Multilateral Agreement.
      We thank the anonymous referee for their careful review and constructive comments, which helped improve the clarity and quality of this manuscript.
\end{acknowledgements}

\bibliographystyle{bibtex/aa} 
\bibliography{aa61416-26}

@ARTICLE{Panuzzo_2024,
       author = {{Gaia Collaboration} and {Panuzzo}, P. and {Mazeh}, T. and {Arenou}, F. and {Holl}, B. and {Caffau}, E. and {Jorissen}, A. and {Babusiaux}, C. and {Gavras}, P. and {Sahlmann}, J. and et al.},
        title = "{Discovery of a dormant 33 solar-mass black hole in pre-release Gaia astrometry}",
      journal = {\aap},
         year = 2024,
        month = jun,
       volume = {686},
          eid = {L2},
        pages = {L2},
          doi = {10.1051/0004-6361/202449763},
archivePrefix = {arXiv},
       eprint = {2404.10486},
 primaryClass = {astro-ph.GA},
       adsurl = {https://ui.adsabs.harvard.edu/abs/2024A&A...686L...2G}
}

@ARTICLE{kepmodel,
       author = {{Delisle}, J.-B. and {S{\'e}gransan}, D.},
        title = "{Analytical determination of orbital elements using Fourier analysis. II. Gaia astrometry and its combination with radial velocities}",
      journal = {\aap},
         year = 2022,
        month = nov,
       volume = {667},
          eid = {A172},
        pages = {A172},
          doi = {10.1051/0004-6361/202244018},
archivePrefix = {arXiv},
       eprint = {2209.13992},
 primaryClass = {astro-ph.EP},
       adsurl = {https://ui.adsabs.harvard.edu/abs/2022A&A...667A.172D}
}

@ARTICLE{Halbwachs_2023,
       author = {{Halbwachs}, Jean-Louis and {Pourbaix}, Dimitri and {Arenou}, Fr{\'e}d{\'e}ric and {Galluccio}, Laurent and {Guillout}, Patrick and {Bauchet}, Nathalie and {Marchal}, Olivier and {Sadowski}, Gilles and {Teyssier}, David},
        title = "{Gaia Data Release 3. Astrometric binary star processing}",
      journal = {\aap},
         year = 2023,
        month = jun,
       volume = {674},
          eid = {A9},
        pages = {A9},
          doi = {10.1051/0004-6361/202243969},
archivePrefix = {arXiv},
       eprint = {2206.05726},
 primaryClass = {astro-ph.SR},
       adsurl = {https://ui.adsabs.harvard.edu/abs/2023A&A...674A...9H}
}

@ARTICLE{DR3,
       author = {{Gaia Collaboration} and {Vallenari}, A. and {Brown}, A.~G.~A. and {Prusti}, T. and {de Bruijne}, J.~H.~J. and {Arenou}, F. and {Babusiaux}, C. and {Biermann}, M. and {Creevey}, O.~L. and {Ducourant}, C. and et al.},
        title = "{Gaia Data Release 3. Summary of the content and survey properties}",
      journal = {\aap},
         year = 2023,
        month = jun,
       volume = {674},
          eid = {A1},
        pages = {A1},
          doi = {10.1051/0004-6361/202243940},
archivePrefix = {arXiv},
       eprint = {2208.00211},
 primaryClass = {astro-ph.GA},
       adsurl = {https://ui.adsabs.harvard.edu/abs/2023A&A...674A...1G}
}

@ARTICLE{Arenou_2023,
       author = {{Gaia Collaboration} and {Arenou}, F. and {Babusiaux}, C. and {Barstow}, M.~A. and {Faigler}, S. and {Jorissen}, A. and {Kervella}, P. and {Mazeh}, T. and {Mowlavi}, N. and {Panuzzo}, P. and et al.},
        title = "{Gaia Data Release 3. Stellar multiplicity, a teaser for the hidden treasure}",
      journal = {\aap},
         year = 2023,
        month = jun,
       volume = {674},
          eid = {A34},
        pages = {A34},
          doi = {10.1051/0004-6361/202243782},
archivePrefix = {arXiv},
       eprint = {2206.05595},
 primaryClass = {astro-ph.SR},
       adsurl = {https://ui.adsabs.harvard.edu/abs/2023A&A...674A..34G}
}

@ARTICLE{Gaia,
       author = {{Gaia Collaboration} and {Prusti}, T. and {de Bruijne}, J.~H.~J. and {Brown}, A.~G.~A. and {Vallenari}, A. and {Babusiaux}, C. and {Bailer-Jones}, C.~A.~L. and {Bastian}, U. and {Biermann}, M. and {Evans}, D.~W. and et al.},
        title = "{The Gaia mission}",
      journal = {\aap},
         year = 2016,
        month = nov,
       volume = {595},
          eid = {A1},
        pages = {A1},
          doi = {10.1051/0004-6361/201629272},
archivePrefix = {arXiv},
       eprint = {1609.04153},
 primaryClass = {astro-ph.IM},
       adsurl = {https://ui.adsabs.harvard.edu/abs/2016A&A...595A...1G}
}

@ARTICLE{Wielen_1996,
       author = {{Wielen}, R.},
        title = "{Searching for VIMs: an astrometric method to detect the binary nature of double stars with a variable component}",
      journal = {\aap},
         year = 1996,
        month = oct,
       volume = {314},
        pages = {679},
       adsurl = {https://ui.adsabs.harvard.edu/abs/1996A&A...314..679W}
}

@ARTICLE{Gallenne_2025,
       author = {{Gallenne}, A. and {Evans}, N.~R. and {Kervella}, P. and {Monnier}, J.~D. and {Proffitt}, C.~R. and {Schaefer}, G.~H. and {Winston}, E.~M. and {Kuraszkiewicz}, J. and {M{\'e}rand}, A. and {Pietrzy{\'n}ski}, G. and et al.},
        title = "{Multiplicity of Galactic Cepheids from long-baseline interferometry: V. High-accuracy orbital parallax and mass of SU Cygni}",
      journal = {\aap},
         year = 2025,
        month = jan,
       volume = {693},
          eid = {A111},
        pages = {A111},
          doi = {10.1051/0004-6361/202452346},
archivePrefix = {arXiv},
       eprint = {2411.06647},
 primaryClass = {astro-ph.SR},
       adsurl = {https://ui.adsabs.harvard.edu/abs/2025A&A...693A.111G}
}

@ARTICLE{Gallenne_2018,
       author = {{Gallenne}, A. and {Kervella}, P. and {Evans}, N.~R. and {Proffitt}, C.~R. and {Monnier}, J.~D. and {M{\'e}rand}, A. and {Nelan}, E. and {Winston}, E. and {Pietrzy{\'n}ski}, G. and {Schaefer}, G. and et al.},
        title = "{A Geometrical 1\% Distance to the Short-period Binary Cepheid V1334 Cygni}",
      journal = {\apj},
         year = 2018,
        month = nov,
       volume = {867},
       number = {2},
          eid = {121},
        pages = {121},
          doi = {10.3847/1538-4357/aae373},
archivePrefix = {arXiv},
       eprint = {1809.07486},
 primaryClass = {astro-ph.SR},
       adsurl = {https://ui.adsabs.harvard.edu/abs/2018ApJ...867..121G}
}

@ARTICLE{Evans_2025,
       author = {{Evans}, Nancy Remage and {Gallenne}, Alexandre and {Kervella}, Pierre and {G{\"u}nther}, H. Moritz and {Kuraszkiewicz}, Joanna and {Anderson}, Richard I. and {Proffitt}, Charles and {M{\'e}rand}, Antoine and {Trahin}, Boris and {Viviani}, Giordano and et al.},
        title = "{Mass of Cepheid V350 Sgr Incorporating Interferometry and the Companion Mass}",
      journal = {\apj},
         year = 2025,
        month = jun,
       volume = {986},
       number = {1},
          eid = {96},
        pages = {96},
          doi = {10.3847/1538-4357/add7dc},
archivePrefix = {arXiv},
       eprint = {2505.13627},
 primaryClass = {astro-ph.SR},
       adsurl = {https://ui.adsabs.harvard.edu/abs/2025ApJ...986...96E}
}

@misc{DR3var,
       author = {{Gaia Collaboration}},
        title = "{VizieR Online Data Catalog: Gaia DR3 Part 4. Variability (Gaia Collaboration, 2022)}",
 howpublished = {VizieR On-line Data Catalog: I/358.  Originally published in: 2023A\&A...674A...1G},
         year = 2022,
        month = may,
          eid = {I/358},
       adsurl = {https://ui.adsabs.harvard.edu/abs/2022yCat.1358....0G}
}

@ARTICLE{Evans_1995,
       author = {{Evans}, Nancy Remage},
        title = "{The Mass Ratios of Cepheid Binaries}",
      journal = {\apj},
         year = 1995,
        month = may,
       volume = {445},
        pages = {393},
          doi = {10.1086/175704},
       adsurl = {https://ui.adsabs.harvard.edu/abs/1995ApJ...445..393E}
}

@ARTICLE{Kervella_2019b,
       author = {{Kervella}, Pierre and {Gallenne}, Alexandre and {Evans}, Nancy Remage and {Szabados}, Laszlo and {Arenou}, Fr{\'e}d{\'e}ric and {M{\'e}rand}, Antoine and {Nardetto}, Nicolas and {Gieren}, Wolfgang and {Pietrzynski}, Grzegorz},
        title = "{Multiplicity of Galactic Cepheids and RR Lyrae stars from Gaia DR2. II. Resolved common proper motion pairs}",
      journal = {\aap},
         year = 2019,
        month = mar,
       volume = {623},
          eid = {A117},
        pages = {A117},
          doi = {10.1051/0004-6361/201834211},
archivePrefix = {arXiv},
       eprint = {1908.00545},
 primaryClass = {astro-ph.SR},
       adsurl = {https://ui.adsabs.harvard.edu/abs/2019A&A...623A.117K}
}

@ARTICLE{Evans_2024,
       author = {{Evans}, Nancy Remage and {Gallenne}, Alexandre and {Kervella}, Pierre and {M{\'e}rand}, Antoine and {Monnier}, John and {Anderson}, Richard I. and {G{\"u}nther}, H. Moritz and {Proffitt}, Charles and {Winston}, Elaine M. and {Pietrzynski}, Grzegorz and et al.},
        title = "{The Orbit and Mass of the Cepheid AW Per}",
      journal = {\apj},
         year = 2024,
        month = sep,
       volume = {972},
       number = {2},
          eid = {145},
        pages = {145},
          doi = {10.3847/1538-4357/ad5e7d},
archivePrefix = {arXiv},
       eprint = {2406.17881},
 primaryClass = {astro-ph.SR},
       adsurl = {https://ui.adsabs.harvard.edu/abs/2024ApJ...972..145E}
}

@ARTICLE{Gallenne_2016,
       author = {{Gallenne}, A. and {Pietrzy{\'n}ski}, G. and {Graczyk}, D. and {Konorski}, P. and {Kervella}, P. and {M{\'e}rand}, A. and {Gieren}, W. and {Anderson}, R.~I. and {Villanova}, S.},
        title = "{The Araucaria Project: High-precision orbital parallax and masses of the eclipsing binary TZ Fornacis}",
      journal = {\aap},
         year = 2016,
        month = feb,
       volume = {586},
          eid = {A35},
        pages = {A35},
          doi = {10.1051/0004-6361/201526764},
archivePrefix = {arXiv},
       eprint = {1511.07971},
 primaryClass = {astro-ph.SR},
       adsurl = {https://ui.adsabs.harvard.edu/abs/2016A&A...586A..35G}
}

@ARTICLE{Gallenne_2023,
       author = {{Gallenne}, A. and {M{\'e}rand}, A. and {Kervella}, P. and {Graczyk}, D. and {Pietrzy{\'n}ski}, G. and {Gieren}, W. and {Pilecki}, B.},
        title = "{The Araucaria project: High-precision orbital parallaxes and masses of binary stars. I. VLTI/GRAVITY observations of ten double-lined spectroscopic binaries}",
      journal = {\aap},
         year = 2023,
        month = apr,
       volume = {672},
          eid = {A119},
        pages = {A119},
          doi = {10.1051/0004-6361/202245712},
archivePrefix = {arXiv},
       eprint = {2302.12960},
 primaryClass = {astro-ph.SR},
       adsurl = {https://ui.adsabs.harvard.edu/abs/2023A&A...672A.119G}
}

@ARTICLE{Szabados_1992,
       author = {{Szabados}, L.},
        title = "{RX Cam is a Spectroscopic-Binary Cepheid}",
      journal = {Information Bulletin on Variable Stars},
         year = 1992,
        month = nov,
       volume = {3812},
        pages = {1},
       adsurl = {https://ui.adsabs.harvard.edu/abs/1992IBVS.3812....1S}
}

@ARTICLE{Shetye_2024,
       author = {{Shetye}, Shreeya S. and {Viviani}, Giordano and {Anderson}, Richard I. and {Mowlavi}, Nami and {Eyer}, Laurent and {Evans}, Nancy R. and {Szabados}, L{\'a}szl{\'o}},
        title = "{VELOcities of CEpheids (VELOCE): II. Systematic search for spectroscopic binary cepheids}",
      journal = {\aap},
         year = 2024,
        month = oct,
       volume = {690},
          eid = {A284},
        pages = {A284},
          doi = {10.1051/0004-6361/202450185},
archivePrefix = {arXiv},
       eprint = {2405.19840},
 primaryClass = {astro-ph.SR},
       adsurl = {https://ui.adsabs.harvard.edu/abs/2024A&A...690A.284S}
}

@INPROCEEDINGS{Kurucz_2003,
       author = {{Castelli}, F. and {Kurucz}, R.~L.},
        title = "{New Grids of ATLAS9 Model Atmospheres}",
    booktitle = {Modelling of Stellar Atmospheres},
         year = 2003,
       editor = {{Piskunov}, N. and {Weiss}, W.~W. and {Gray}, D.~F.},
       series = {IAU Symposium},
       volume = {210},
        month = jan,
        pages = {A20},
          doi = {10.48550/arXiv.astro-ph/0405087},
archivePrefix = {arXiv},
       eprint = {astro-ph/0405087},
 primaryClass = {astro-ph},
       adsurl = {https://ui.adsabs.harvard.edu/abs/2003IAUS..210P.A20C}
}

@ARTICLE{Gallenne_2019,
       author = {{Gallenne}, A. and {Kervella}, P. and {Borgniet}, S. and {M{\'e}rand}, A. and {Pietrzy{\'n}ski}, G. and {Gieren}, W. and {Monnier}, J.~D. and {Schaefer}, G.~H. and {Evans}, N.~R. and {Anderson}, R.~I. and et al.},
        title = "{Multiplicity of Galactic Cepheids from long-baseline interferometry. IV. New detected companions from MIRC and PIONIER observations}",
      journal = {\aap},
         year = 2019,
        month = feb,
       volume = {622},
          eid = {A164},
        pages = {A164},
          doi = {10.1051/0004-6361/201834614},
archivePrefix = {arXiv},
       eprint = {1812.09989},
 primaryClass = {astro-ph.SR},
       adsurl = {https://ui.adsabs.harvard.edu/abs/2019A&A...622A.164G}
}

@ARTICLE{Kervella_2022,
       author = {{Kervella}, Pierre and {Arenou}, Fr{\'e}d{\'e}ric and {Th{\'e}venin}, Fr{\'e}d{\'e}ric},
        title = "{Stellar and substellar companions from Gaia EDR3. Proper-motion anomaly and resolved common proper-motion pairs}",
      journal = {\aap},
         year = 2022,
        month = jan,
       volume = {657},
          eid = {A7},
        pages = {A7},
          doi = {10.1051/0004-6361/202142146},
archivePrefix = {arXiv},
       eprint = {2109.10912},
 primaryClass = {astro-ph.SR},
       adsurl = {https://ui.adsabs.harvard.edu/abs/2022A&A...657A...7K}
}

@ARTICLE{Lindegren_2021,
       author = {{Lindegren}, L. and {Bastian}, U. and {Biermann}, M. and {Bombrun}, A. and {de Torres}, A. and {Gerlach}, E. and {Geyer}, R. and {Hern{\'a}ndez}, J. and {Hilger}, T. and {Hobbs}, D. and {Klioner}, S.~A. and {Lammers}, U. and {McMillan}, P.~J. and {Ramos-Lerate}, M. and {Steidelm{\"u}ller}, H. and {Stephenson}, C.~A. and {van Leeuwen}, F.},
        title = "{Gaia Early Data Release 3. Parallax bias versus magnitude, colour, and position}",
      journal = {\aap},
         year = 2021,
        month = may,
       volume = {649},
          eid = {A4},
        pages = {A4},
          doi = {10.1051/0004-6361/202039653},
archivePrefix = {arXiv},
       eprint = {2012.01742},
 primaryClass = {astro-ph.IM},
       adsurl = {https://ui.adsabs.harvard.edu/abs/2021A&A...649A...4L}
}

@ARTICLE{Storm_2011,
       author = {{Storm}, J. and {Gieren}, W. and {Fouqu{\'e}}, P. and {Barnes}, T.~G. and {Pietrzy{\'n}ski}, G. and {Nardetto}, N. and {Weber}, M. and {Granzer}, T. and {Strassmeier}, K.~G.},
        title = "{Calibrating the Cepheid period-luminosity relation from the infrared surface brightness technique. I. The p-factor, the Milky Way relations, and a universal K-band relation}",
      journal = {\aap},
         year = 2011,
        month = oct,
       volume = {534},
          eid = {A94},
        pages = {A94},
          doi = {10.1051/0004-6361/201117155},
archivePrefix = {arXiv},
       eprint = {1109.2017},
 primaryClass = {astro-ph.CO},
       adsurl = {https://ui.adsabs.harvard.edu/abs/2011A&A...534A..94S}
}

@ARTICLE{Mamajek_2013,
       author = {{Pecaut}, Mark J. and {Mamajek}, Eric E.},
        title = "{Intrinsic Colors, Temperatures, and Bolometric Corrections of Pre-main-sequence Stars}",
      journal = {\apjs},
         year = 2013,
        month = sep,
       volume = {208},
       number = {1},
          eid = {9},
        pages = {9},
          doi = {10.1088/0067-0049/208/1/9},
archivePrefix = {arXiv},
       eprint = {1307.2657},
 primaryClass = {astro-ph.SR},
       adsurl = {https://ui.adsabs.harvard.edu/abs/2013ApJS..208....9P}
}

@ARTICLE{Moe2017ApJS..230...15M,
       author = {{Moe}, Maxwell and {Di Stefano}, Rosanne},
        title = "{Mind Your Ps and Qs: The Interrelation between Period (P) and Mass-ratio (Q) Distributions of Binary Stars}",
      journal = {\apjs},
         year = 2017,
        month = jun,
       volume = {230},
       number = {2},
          eid = {15},
        pages = {15},
          doi = {10.3847/1538-4365/aa6fb6},
archivePrefix = {arXiv},
       eprint = {1606.05347},
 primaryClass = {astro-ph.SR},
       adsurl = {https://ui.adsabs.harvard.edu/abs/2017ApJS..230...15M}
}

@ARTICLE{Fulton2018AJ....156..264F,
       author = {{Fulton}, Benjamin J. and {Petigura}, Erik A.},
        title = "{The California-Kepler Survey. VII. Precise Planet Radii Leveraging Gaia DR2 Reveal the Stellar Mass Dependence of the Planet Radius Gap}",
      journal = {\aj},
         year = 2018,
        month = dec,
       volume = {156},
       number = {6},
          eid = {264},
        pages = {264},
          doi = {10.3847/1538-3881/aae828},
archivePrefix = {arXiv},
       eprint = {1805.01453},
 primaryClass = {astro-ph.EP},
       adsurl = {https://ui.adsabs.harvard.edu/abs/2018AJ....156..264F}
}

@INPROCEEDINGS{Offner2023ASPC..534..275O,
       author = {{Offner}, S.~S.~R. and {Moe}, M. and {Kratter}, K.~M. and {Sadavoy}, S.~I. and {Jensen}, E.~L.~N. and {Tobin}, J.~J.},
        title = "{The Origin and Evolution of Multiple Star Systems}",
    booktitle = {Protostars and Planets VII},
         year = 2023,
       editor = {{Inutsuka}, S. and {Aikawa}, Y. and {Muto}, T. and {Tomida}, K. and {Tamura}, M.},
       series = {Astronomical Society of the Pacific Conference Series},
       volume = {534},
        month = jul,
        pages = {275},
          doi = {10.48550/arXiv.2203.10066},
archivePrefix = {arXiv},
       eprint = {2203.10066},
 primaryClass = {astro-ph.SR},
       adsurl = {https://ui.adsabs.harvard.edu/abs/2023ASPC..534..275O}
}

@ARTICLE{H0DN,
       author = {{H0DN Collaboration} and {Casertano}, Stefano and {Anand}, Gagandeep and {Anderson}, Richard I. and {Beaton}, Rachael and {Bhardwaj}, Anupam and {Blakeslee}, John P. and {Boubel}, Paula and {Breuval}, Louise and {Brout}, Dillon and {Cantiello}, Michele and {Cruz Reyes}, Mauricio and {Cs{\"o}rnyei}, Geza and {de Jaeger}, Thomas and {Dhawan}, Suhail and {Di Valentino}, Eleonora and {Galbany}, Llu{\'\i}s and {Gil-Mar{\'\i}n}, H{\'e}ctor and {Graczyk}, Dariusz and {Huang}, Caroline and {Jensen}, Joseph B. and {Kervella}, Pierre and {Leibundgut}, Bruno and {Lengen}, Bastian and {Li}, Siyang and {Macri}, Lucas and {{\"O}z{\"u}lker}, Emre and {Pesce}, Dominic W. and {Riess}, Adam and {Romaniello}, Martino and {Said}, Khaled and {Sch{\"o}neberg}, Nils and {Scolnic}, Dan and {Sicignano}, Teresa and {Skowron}, Dorota M. and {Uddin}, Syed A. and {Verde}, Licia and {Nota}, Antonella},
        title = "{The Local Distance Network: A community consensus report on the measurement of the Hubble constant at {\ensuremath{\sim}}1\% precision}",
      journal = {\aap},
         year = 2026,
        month = apr,
       volume = {708},
          eid = {A166},
        pages = {A166},
          doi = {10.1051/0004-6361/202557993},
archivePrefix = {arXiv},
       eprint = {2510.23823},
 primaryClass = {astro-ph.CO},
       adsurl = {https://ui.adsabs.harvard.edu/abs/2026A&A...708A.166H}
}

@misc{Brown_2025,
  author       = {Brown, Anthony George Alxeander},
  title        = {Gaia DR4 Processing Status and Contents Preview},
  month        = sep,
  year         = 2025,
  publisher    = {Zenodo},
  doi          = {10.5281/zenodo.17158744},
  url          = {https://doi.org/10.5281/zenodo.17158744},
}

@article{Leavitt_1908,
  title={1777 variables in the Magellanic Clouds},
  author={Leavitt, Henrietta S},
  journal={Annals of Harvard College Observatory, vol. 60, pp. 87-108.3},
  volume={60},
  pages={87--108},
  year={1908}
}

@ARTICLE{Leavitt_1912,
       author = {{Leavitt}, Henrietta S. and {Pickering}, Edward C.},
        title = "{Periods of 25 Variable Stars in the Small Magellanic Cloud.}",
      journal = {Harvard College Observatory Circular},
         year = 1912,
        month = mar,
       volume = {173},
        pages = {1-3},
       adsurl = {https://ui.adsabs.harvard.edu/abs/1912HarCi.173....1L}
}

@INPROCEEDINGS{Sana_2016,
       author = {{Sana}, Hugues},
        title = "{The multiplicity of massive stars: a 2016 view}",
    booktitle = {The Lives and Death-Throes of Massive Stars},
         year = 2017,
       editor = {{Eldridge}, J.~J. and {Bray}, J.~C. and {McClelland}, L.~A.~S. and {Xiao}, L.},
       series = {IAU Symposium},
       volume = {329},
        month = nov,
        pages = {110-117},
          doi = {10.1017/S1743921317003209},
archivePrefix = {arXiv},
       eprint = {1703.01608},
 primaryClass = {astro-ph.SR},
       adsurl = {https://ui.adsabs.harvard.edu/abs/2017IAUS..329..110S}
}

@ARTICLE{Evans_2015,
       author = {{Evans}, Nancy Remage and {Berdnikov}, Leonid and {Lauer}, Jennifer and {Morgan}, Douglas and {Nichols}, Joy and {G{\"u}nther}, H. Moritz and {Gorynya}, Natalya and {Rastorguev}, Alexey and {Moskalik}, Pawel},
        title = "{Binary Properties from Cepheid Radial Velocities (CRaV)}",
      journal = {\aj},
         year = 2015,
        month = jul,
       volume = {150},
       number = {1},
          eid = {13},
        pages = {13},
          doi = {10.1088/0004-6256/150/1/13},
archivePrefix = {arXiv},
       eprint = {1505.05823},
 primaryClass = {astro-ph.SR},
       adsurl = {https://ui.adsabs.harvard.edu/abs/2015AJ....150...13E}
}

@ARTICLE{Kervella_2019a,
       author = {{Kervella}, Pierre and {Gallenne}, Alexandre and {Remage Evans}, Nancy and {Szabados}, Laszlo and {Arenou}, Fr{\'e}d{\'e}ric and {M{\'e}rand}, Antoine and {Proto}, Yann and {Karczmarek}, Paulina and {Nardetto}, Nicolas and {Gieren}, Wolfgang and et al.},
        title = "{Multiplicity of Galactic Cepheids and RR Lyrae stars from Gaia DR2. I. Binarity from proper motion anomaly}",
      journal = {\aap},
         year = 2019,
        month = mar,
       volume = {623},
          eid = {A116},
        pages = {A116},
          doi = {10.1051/0004-6361/201834210},
archivePrefix = {arXiv},
       eprint = {1903.03632},
 primaryClass = {astro-ph.SR},
       adsurl = {https://ui.adsabs.harvard.edu/abs/2019A&A...623A.116K}
}

@ARTICLE{Beguin_2024,
       author = {{B{\'e}guin}, E. and {Chiavassa}, A. and {Ahmad}, A. and {Freytag}, B. and {Uttenthaler}, S.},
        title = "{Retrieving stellar parameters and dynamics of AGB stars with Gaia parallax measurements and CO$^{5}$BOLD RHD simulations}",
      journal = {\aap},
         year = 2024,
        month = oct,
       volume = {690},
          eid = {A125},
        pages = {A125},
          doi = {10.1051/0004-6361/202450245},
archivePrefix = {arXiv},
       eprint = {2409.03422},
 primaryClass = {astro-ph.SR},
       adsurl = {https://ui.adsabs.harvard.edu/abs/2024A&A...690A.125B}
}

@ARTICLE{Lindegren_2012,
       author = {{Lindegren}, L. and {Lammers}, U. and {Hobbs}, D. and {O'Mullane}, W. and {Bastian}, U. and {Hern{\'a}ndez}, J.},
        title = "{The astrometric core solution for the Gaia mission. Overview of models, algorithms, and software implementation}",
      journal = {\aap},
         year = 2012,
        month = feb,
       volume = {538},
          eid = {A78},
        pages = {A78},
          doi = {10.1051/0004-6361/201117905},
archivePrefix = {arXiv},
       eprint = {1112.4139},
 primaryClass = {astro-ph.IM},
       adsurl = {https://ui.adsabs.harvard.edu/abs/2012A&A...538A..78L}
}

@ARTICLE{ElBadry2024OJAp....7E.100E,
       author = {{El-Badry}, Kareem and {Lam}, Casey and {Holl}, Berry and {Halbwachs}, Jean-Louis and {Rix}, Hans-Walter and {Mazeh}, Tsevi and {Shahaf}, Sahar},
        title = "{A generative model for Gaia astrometric orbit catalogs: selection functions for binary stars, giant planets, and compact object companions}",
      journal = {The Open Journal of Astrophysics},
         year = 2024,
        month = nov,
       volume = {7},
          eid = {100},
        pages = {100},
          doi = {10.33232/001c.125461},
archivePrefix = {arXiv},
       eprint = {2411.00088},
 primaryClass = {astro-ph.SR},
       adsurl = {https://ui.adsabs.harvard.edu/abs/2024OJAp....7E.100E}
}

@ARTICLE{Eyer_2023,
       author = {{Eyer}, L. and {Audard}, M. and {Holl}, B. and {Rimoldini}, L. and {Carnerero}, M.~I. and {Clementini}, G. and {De Ridder}, J. and {Distefano}, E. and {Evans}, D.~W. and {Gavras}, P. and {Gomel}, R. and {Lebzelter}, T. and {Marton}, G. and {Mowlavi}, N. and {Panahi}, A. and {Ripepi}, V. and {Wyrzykowski}, {\L}. and {Nienartowicz}, K. and {Jevardat de Fombelle}, G. and {Lecoeur-Taibi}, I. and {Rohrbasser}, L. and {Riello}, M. and {Garc{\'\i}a-Lario}, P. and {Lanzafame}, A.~C. and {Mazeh}, T. and {Raiteri}, C.~M. and {Zucker}, S. and {{\'A}brah{\'a}m}, P. and {Aerts}, C. and {Aguado}, J.~J. and {Anderson}, R.~I. and {Bashi}, D. and {Binnenfeld}, A. and {Faigler}, S. and {Garofalo}, A. and {Karbevska}, L. and {K{\'o}sp{\'a}l}, {\'A}. and {Kruszy{\'n}ska}, K. and {Kun}, M. and {Lanza}, A.~F. and {Leccia}, S. and {Marconi}, M. and {Messina}, S. and {Molinaro}, R. and {Moln{\'a}r}, L. and {Muraveva}, T. and {Musella}, I. and {Nagy}, Z. and {Pagano}, I. and {Palaversa}, L. and {Plachy}, E. and {Pr{\v{s}}a}, A. and {Rybicki}, K.~A. and {Shahaf}, S. and {Szabados}, L. and {Szegedi-Elek}, E. and {Trabucchi}, M. and {Barblan}, F. and {Grenon}, M. and {Roelens}, M. and {S{\"u}veges}, M.},
        title = "{Gaia Data Release 3. Summary of the variability processing and analysis}",
      journal = {\aap},
         year = 2023,
        month = jun,
       volume = {674},
          eid = {A13},
        pages = {A13},
          doi = {10.1051/0004-6361/202244242},
archivePrefix = {arXiv},
       eprint = {2206.06416},
 primaryClass = {astro-ph.SR},
       adsurl = {https://ui.adsabs.harvard.edu/abs/2023A&A...674A..13E}
}

@ARTICLE{2026arXiv260514128S,
       author = {{Smolec}, R. and {Singh Rathour}, R. and {Hocd{\'e}}, V.},
        title = "{Toward a Comprehensive Grid of Cepheid Models with MESA. IV. Modest Effects of Rotation on Blue Loops}",
      journal = {arXiv e-prints},
         year = 2026,
        month = may,
          eid = {arXiv:2605.14128},
        pages = {arXiv:2605.14128},
archivePrefix = {arXiv},
       eprint = {2605.14128},
 primaryClass = {astro-ph.SR},
       adsurl = {https://ui.adsabs.harvard.edu/abs/2026arXiv260514128S}
}

@ARTICLE{Neilson_2011,
       author = {{Neilson}, H.~R. and {Cantiello}, M. and {Langer}, N.},
        title = "{The Cepheid mass discrepancy and pulsation-driven mass loss}",
      journal = {\aap},
         year = 2011,
        month = may,
       volume = {529},
          eid = {L9},
        pages = {L9},
          doi = {10.1051/0004-6361/201116920},
archivePrefix = {arXiv},
       eprint = {1104.1638},
 primaryClass = {astro-ph.SR},
       adsurl = {https://ui.adsabs.harvard.edu/abs/2011A&A...529L...9N}
}

@ARTICLE{Groenewegen_2013,
       author = {{Groenewegen}, M.~A.~T.},
        title = "{Baade-Wesselink distances to Galactic and Magellanic Cloud Cepheids and the effect of metallicity}",
      journal = {\aap},
         year = 2013,
        month = feb,
       volume = {550},
          eid = {A70},
        pages = {A70},
          doi = {10.1051/0004-6361/201220446},
archivePrefix = {arXiv},
       eprint = {1212.5478},
 primaryClass = {astro-ph.GA},
       adsurl = {https://ui.adsabs.harvard.edu/abs/2013A&A...550A..70G}
}

@ARTICLE{Nardetto_2023,
       author = {{Nardetto}, N. and {Gieren}, W. and {Storm}, J. and {Hocd{\'e}}, V. and {Pietrzy{\'n}ski}, G. and {Kervella}, P. and {M{\'e}rand}, A. and {Gallenne}, A. and {Graczyk}, D. and {Pilecki}, B. and {Poretti}, E. and {Rainer}, M. and {Zgirski}, B. and {Wielg{\'o}rski}, P. and {Hajdu}, G. and {G{\'o}rski}, M. and {Karczmarek}, P. and {Narloch}, W. and {Taormina}, M.},
        title = "{HARPS-N high spectral resolution observations of Cepheids. II. The impact of the surface-brightness color relation on the Baade-Wesselink projection factor of {\ensuremath{\eta}} Aql}",
      journal = {\aap},
         year = 2023,
        month = mar,
       volume = {671},
          eid = {A14},
        pages = {A14},
          doi = {10.1051/0004-6361/202245298},
archivePrefix = {arXiv},
       eprint = {2301.09468},
 primaryClass = {astro-ph.SR},
       adsurl = {https://ui.adsabs.harvard.edu/abs/2023A&A...671A..14N}
}

@ARTICLE{Riess_2022,
       author = {{Riess}, Adam G. and {Yuan}, Wenlong and {Macri}, Lucas M. and {Scolnic}, Dan and {Brout}, Dillon and {Casertano}, Stefano and {Jones}, David O. and {Murakami}, Yukei and {Anand}, Gagandeep S. and {Breuval}, Louise and {Brink}, Thomas G. and {Filippenko}, Alexei V. and {Hoffmann}, Samantha and {Jha}, Saurabh W. and {D'arcy Kenworthy}, W. and {Mackenty}, John and {Stahl}, Benjamin E. and {Zheng}, WeiKang},
        title = "{A Comprehensive Measurement of the Local Value of the Hubble Constant with 1 km s$^{-1}$ Mpc$^{-1}$ Uncertainty from the Hubble Space Telescope and the SH0ES Team}",
      journal = {\apjl},
         year = 2022,
        month = jul,
       volume = {934},
       number = {1},
          eid = {L7},
        pages = {L7},
          doi = {10.3847/2041-8213/ac5c5b},
archivePrefix = {arXiv},
       eprint = {2112.04510},
 primaryClass = {astro-ph.CO},
       adsurl = {https://ui.adsabs.harvard.edu/abs/2022ApJ...934L...7R}
}

@ARTICLE{Breuval_2024,
       author = {{Breuval}, Louise and {Riess}, Adam G. and {Casertano}, Stefano and {Yuan}, Wenlong and {Macri}, Lucas M. and {Romaniello}, Martino and {Murakami}, Yukei S. and {Scolnic}, Daniel and {Anand}, Gagandeep S. and {Soszy{\'n}ski}, Igor},
        title = "{Small Magellanic Cloud Cepheids Observed with the Hubble Space Telescope Provide a New Anchor for the SH0ES Distance Ladder}",
      journal = {\apj},
         year = 2024,
        month = sep,
       volume = {973},
       number = {1},
          eid = {30},
        pages = {30},
          doi = {10.3847/1538-4357/ad630e},
archivePrefix = {arXiv},
       eprint = {2404.08038},
 primaryClass = {astro-ph.CO},
       adsurl = {https://ui.adsabs.harvard.edu/abs/2024ApJ...973...30B}
}

@ARTICLE{2023A&A...674A..10H,
       author = {{Holl}, B. and {Sozzetti}, A. and {Sahlmann}, J. and {Giacobbe}, P. and {S{\'e}gransan}, D. and {Unger}, N. and {Delisle}, J.-B. and {Barbato}, D. and {Lattanzi}, M.~G. and {Morbidelli}, R. and {Sosnowska}, D.},
        title = "{Gaia Data Release 3. Astrometric orbit determination with Markov chain Monte Carlo and genetic algorithms: Systems with stellar, sub-stellar, and planetary mass companions}",
      journal = {\aap},
         year = 2023,
        month = jun,
       volume = {674},
          eid = {A10},
        pages = {A10},
          doi = {10.1051/0004-6361/202244161},
archivePrefix = {arXiv},
       eprint = {2206.05439},
 primaryClass = {astro-ph.EP},
       adsurl = {https://ui.adsabs.harvard.edu/abs/2023A&A...674A..10H}
}

@ARTICLE{Decin_2026,
       author = {{Decin}, L. and {Sivkova}, K. and {Kervella}, P. and {Chiavassa}, A. and {B{\'e}guin}, E.},
        title = "{Astrometric modelling of unresolved variable binary systems. II. Application to Gaia epoch astrometry of nearby pulsating and convective red giants}",
      journal = {\aap},
         year = {2026},
        month = aug,
      note    = {in press},
      doi     = {10.1051/0004-6361/202661421}
}

@ARTICLE{Everall_2021,
       author = {{Everall}, Andrew and {Boubert}, Douglas and {Koposov}, Sergey E. and {Smith}, Leigh and {Holl}, Berry},
        title = "{Completeness of the Gaia-verse - IV. The astrometry spread function of Gaia DR2}",
      journal = {\mnras},
         year = 2021,
        month = apr,
       volume = {502},
       number = {2},
        pages = {1908-1924},
          doi = {10.1093/mnras/stab041},
archivePrefix = {arXiv},
       eprint = {2101.01723},
 primaryClass = {astro-ph.GA},
       adsurl = {https://ui.adsabs.harvard.edu/abs/2021MNRAS.502.1908E}
}

\begin{appendix} \label{annex}

\section{VIM models \label{annex:vim}}

In addition to astrometric measurements, VIM models also require simultaneous flux (or magnitude) measurements of the system. Following \citet{Wielen_1996}, we introduce

\begin{equation}
f(t) = \frac{\overline{F}}{F(t)} - 1 
= 10^{0.4\,(m_{\mathrm{tot}}(t) - \overline{m})} - 1,
\label{fref}
\end{equation}

where $F(t) = F_1(t) + F_2$ (or $m_{\mathrm{tot}}(t)$) denotes the total flux (or magnitude) of the system, and the overbar denotes a reference value, in practice chosen as the mean.

\begin{figure}[hbt!]
  \centering \includegraphics[width=.99\hsize]{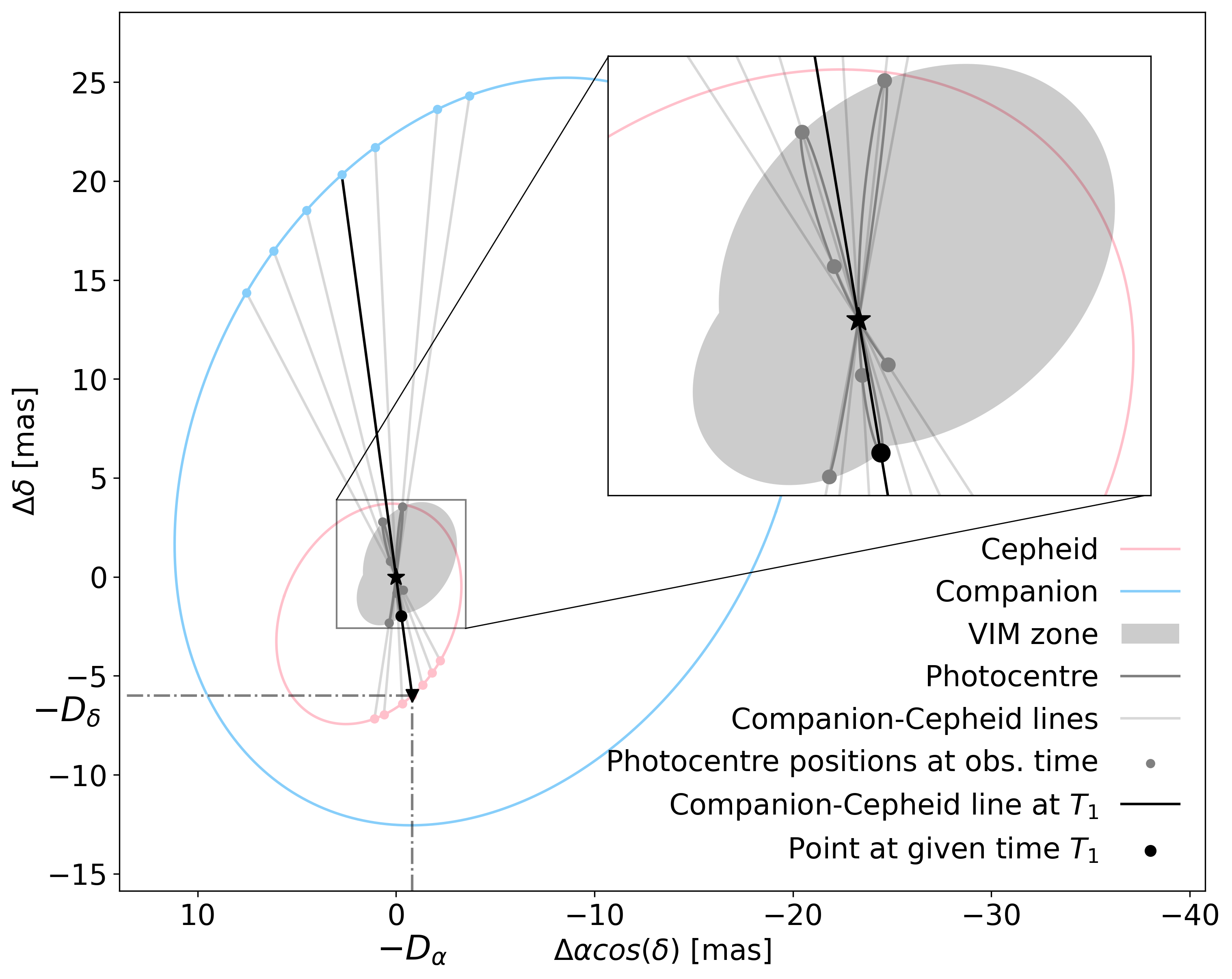}
  \caption{The same VIM system as in Fig.~\ref{VIM_exp} with the VIM effect exaggerated, but shown in a reference frame centred on the mean photocentre position (dashed line in Fig.~\ref{VIM_exp}), corresponding to the reference flux defined in Eq.~\ref{fref}. The photocentre moves along the line connecting the two components, within the VIM zone (grey area). The parameters $D_{\alpha}$ and $D_{\delta}$ denote the coordinates of the photocentre relative to the Cepheid.}
  \label{VIM_exp_phot}
\end{figure}

The VIM fixed (VIMF) model, already used in DR3 \citep{Halbwachs_2023}, extends the single-star model with two additional parameters, $D_{\alpha}$ and $D_{\delta}$ (see Fig.~\ref{VIM_exp_phot}).These parameters represent the link between the astrometric positions and the flux of the system. This model assumes that the two components remain static with respect to each other.

\begin{equation}
\label{vimf}
w_{\mathrm{VIMF}} = w_{\mathrm{ss}} 
+ \left( D_{\alpha} \sin \theta + D_{\delta} \cos \theta \right) f(t).
\end{equation}

The vector $D = \sqrt{D_{\alpha}^{2} + D_{\delta}^{2}}$ indicates the direction from the Cepheid to the companion and with its magnitude being the separation between the Cepheid and the photocentre.

We similarly define the VIM linear (VIML) and accelerated (VIMA) models, which approximate linear and accelerated motion of the components in the centre-of-mass frame:

\begin{equation}
\label{viml}
w_{\mathrm{VIML}} = w_{\mathrm{ss}} 
+ \left( D_{\alpha} + \dot{D}_{\alpha}t \right) f(t)\,\sin \theta 
+ \left( D_{\delta} + \dot{D}_{\delta}t \right) f(t)\,\cos \theta,
\end{equation}

and

\begin{equation}
\begin{aligned}
\label{vima}
w_{\mathrm{VIMA}} &= w_{\mathrm{ss}} \\
&+ \left[ \frac{1}{2}\dot{\mu}_{\alpha} t^2 
+ \left( D_{\alpha} + \dot{D}_{\alpha} t + \frac{1}{2}\ddot{D}_{\alpha} t^2 \right) f(t) \right]\sin \theta \\
&+ \left[ \frac{1}{2}\dot{\mu}_{\delta} t^2 
+ \left( D_{\delta} + \dot{D}_{\delta} t + \frac{1}{2}\ddot{D}_{\delta} t^2 \right) f(t) \right]\cos \theta,
\end{aligned}
\end{equation}

where the overdot denotes the derivative and the double overdot the second derivative. The additional $D$ parameters represent the proper motion and acceleration parameters needed to describe the change in the direction of the VIM effect caused by the orbital motion of the components.

\medskip

In the case of the variability-induced mover orbital (VIMO) model, we express the orbital contribution from Eq.~\eqref{Eq:aph} as

\begin{equation}
\begin{aligned}
\alpha^\star_{\mathrm{orb}}(t) &= \left( \tilde A f(t) - \overline{a_{\mathrm{ph}}} \right) (B X + G Y), \\
\delta_{\mathrm{orb}}(t) &= \left( \tilde A f(t) - \overline{a_{\mathrm{ph}}} \right) (A X + F Y),
\end{aligned}
\end{equation}

where $A$, $B$, $F$, and $G$ are the Thiele--Innes constants, and

\begin{equation}
\begin{aligned}
X &= \cos E - e, \\
Y &= \sqrt{1-e^2}\,\sin E,
\end{aligned}
\end{equation}

with the eccentric anomaly $E$ determined from Kepler's equation,

\begin{equation}
E - e \sin E = \frac{2\pi (t-T_0)}{P}.
\end{equation}

The parameters $\tilde A$ and $\overline{a_{\mathrm{ph}}}$ are specific to the VIMO model and are defined as

\begin{equation}
\begin{aligned}
\tilde A &= \overline{\beta}\, a_{\mathrm{rel}}
   = \frac{F_2}{\overline{F}}\, a_{\mathrm{rel}}, \\
\overline{a_{\mathrm{ph}}} &= (B - \overline{\beta})\, a_{\mathrm{rel}}.
\end{aligned}
\end{equation}

In this case, the parameter $A$ links the astrometric variations to the flux variations, similarly to the ensemble of $D$ parameters.

\section{Fitting procedure and uncertainty estimation \label{annex:fitting}}

The fits were performed following the procedures detailed below for each model. Parameter uncertainties were estimated using a bootstrap method, in which 10\% of the predicted transits and an additional 10\% of the per-CCD measurements were randomly rejected at each iteration.

\subsection{SS, VIMF, and VIML models}

The single-star, VIM fixed-geometry, and VIM linear-motion models are linear in their fitted parameters. Their equations can be written as

\begin{equation} \mathbf{w} = \mathbf{A}\mathbf{p}, \end{equation}

where $\mathbf{p}$ is the parameter vector and $\mathbf{A}$ is the
design matrix. For the VIML model, $\mathbf{p}$ is defined as follows;
see Eq.~\eqref{matrixA} for the definition of $\mathbf{A}$,

\begin{equation}
\mathbf{p} =
\begin{pmatrix}
\Delta \alpha^\star \\
\mu_{\alpha^\star} \\
\Delta \delta \\
\mu_\delta \\
\varpi \\
D_{\alpha^\star} \\
\dot{D}_{\alpha^\star} \\
D_\delta \\
\dot{D}_\delta
\end{pmatrix},
\end{equation}

\begin{figure*}[t]
\begin{equation}
    \label{matrixA}
\mathbf{A} =
\begin{pmatrix}
\sin\theta_1 &
t_1\sin\theta_1 &
\cos\theta_1 &
t_1\cos\theta_1 &
\Pi_{\varpi,1} &
f(t_1)\sin\theta_1 &
t_1f(t_1)\sin\theta_1 &
f(t_1)\cos\theta_1 &
t_1f(t_1)\cos\theta_1
\\
\sin\theta_2 &
t_2\sin\theta_2 &
\cos\theta_2 &
t_2\cos\theta_2 &
\Pi_{\varpi,2} &
f(t_2)\sin\theta_2 &
t_2f(t_2)\sin\theta_2 &
f(t_2)\cos\theta_2 &
t_2f(t_2)\cos\theta_2
\\
\vdots & \vdots & \vdots & \vdots & \vdots &
\vdots & \vdots & \vdots & \vdots
\\
\sin\theta_n &
t_n\sin\theta_n &
\cos\theta_n &
t_n\cos\theta_n &
\Pi_{\varpi,n} &
f(t_n)\sin\theta_n &
t_nf(t_n)\sin\theta_n &
f(t_n)\cos\theta_n &
t_nf(t_n)\cos\theta_n
\end{pmatrix}
\end{equation}
\end{figure*}

while for the SS and VIMF models, they are the same but contain fewer parameters, depending on the model (see Appendix \ref{annex:vim}).

The covariance matrix is constructed from the uncertainties of the along-scan measurements:

\begin{equation}
\mathbf{C}^{-1}
=
\operatorname{diag}
\left(
\frac{1}{w_{\mathrm{err},i}^{2}}
\right)_{i=1,\ldots,n},
\end{equation}

we consider the individual astrometric measurements to be uncorrelated, as explained in Section~\ref{Sec:AL_astro}.

The fitted parameter vector $\mathbf{p}$ is then obtained directly through weighted least squares:

\begin{equation}
\mathbf{p}  =
(\mathbf{A}^{\mathrm T} C^{-1} \mathbf{A})^{-1}
\mathbf{A}^{\mathrm T} C^{-1} \mathbf{w}.
\end{equation}

\subsection{VIMA model}

Unlike the SS, VIMF, and VIML models, the parametrisation of the VIMA model is nonlinear because of the acceleration terms. The acceleration vector $(\ddot{D}_\alpha^\star, \ddot{D}_\delta)$ is required to be anti-parallel to the positional vector $(D_\alpha^\star, D_\delta)$, while the proper-motion variation vector $(\dot{\mu}_{\alpha^\star}, \dot{\mu}_{\delta})$ must be parallel or anti-parallel to $(\ddot{D}_\alpha^\star, \ddot{D}_\delta)$. To model this dependence, we instead introduce two scalar parameters, $k$ and $s$, rather than directly fitting $(\ddot{D}_\alpha^\star, \ddot{D}_\delta)$ and $(\dot{\mu}_{\alpha^\star}, \dot{\mu}_{\delta})$. The corresponding vectors are then derived as:

\begin{equation}
\begin{aligned}
\ddot{D}_\alpha^\star &= -k\, D_\delta \\
\ddot{D}_\delta &= -k\, D_\alpha^\star
\end{aligned}
\end{equation}

\begin{equation}
\begin{aligned}
\dot{\mu}_{\alpha^\star} &= s\, \ddot{D}_\alpha^\star \\
\dot{\mu}_{\delta} &= s\, \ddot{D}_\delta .
\end{aligned}
\end{equation}

Similarly to the $\mathbf{A}$ matrix and $\mathbf{p}$ vector for the linear models, the VIMA model is described by a Jacobian matrix $\mathbf{J}$ and a parameter vector $\mathbf{p}$, which includes the two additional scalar parameters $k$ and $s$. The model parameters are then optimised by nonlinear least-squares minimization of the residuals between the data and the model.

\section{Tables and figures}

\begin{table}
\caption{Computed $a_{\mathrm{ph}}$ for the non-pulsating binary sample.}
\label{tab:binary}
\centering
\begin{tabular}{lccc}
\hline\hline
Name & $f_{K}$ (\%) & $f_{G}$ (\%) & $a_{\mathrm{ph}}$ (mas) \\
\hline
TZ For    & 31.8  & 97.8 & 0.02 \\
AK For    & 71.7  & 55.7 & 0.21 \\
HD 9312   & 7.84  & 6.7  & 1.79 \\
HD 41255  & 95.9  & 98.6 & 0.06 \\
HD 70937  & 59.7  & 65.1 & 0.33 \\
HD 210763 & 37.1  & 44.8 & 0.57 \\
HD 224974 & 90.0  & 87.3 & 0.06 \\
HD 188088 & 90.1  & 90.1 & 0.47 \\
LL Aqr    & 53.3  & 42.3 & 0.23 \\
o~Leo     & 25.03 & 53.9 & 0.54 \\
V963 Cen  & 96.0  & 96.6 & 0.01 \\
\hline
\end{tabular}
\tablefoot{$f_{K}$ are taken from \citet{Gallenne_2023}.}
\end{table}

\begin{table}
\caption{Campbell orbital parameters for AK~For used for the simulation and as listed in the Gaia DR3 catalogue.}
\label{tab:AK For}
\centering
\small
\setlength{\tabcolsep}{4pt}
\begin{tabular}{lcc}
\hline\hline
Parameter & Simulation & Gaia DR3 \\
\hline
$P$ [days]              & $3.9809943 \pm 0.0000004^{a}$ & $3.9785 \pm 0.0006$ \\
$a_{\mathrm{ph}}$ [mas] & $0.21 \pm 0.01$               & $0.20 \pm 0.01$ \\
$e$                     & $0.0002 \pm 0.0002^{a}$       & $0.287 \pm 0.097$ \\
$i$ [deg]               & $86.9 \pm 0.1^{a}$            & $86.6 \pm 3.1$ \\
$\Omega$ [deg]          & $190.0 \pm 0.1^{a}$           & $11.096 \pm 3.4$ \\
$\omega$ [deg]          & $170 \pm 107^{a}$             & $226.5 \pm 17.6$ \\
$T_{0}$ [JD]            & $2451904 \pm 1^{a}$  & $2457389.894 \pm 0.387$ \\
$\varpi$ [mas]          & $32.147 \pm 0.055^{a}$        & $32.158 \pm 0.017$ \\
$\mu_{\alpha^\star}$ [mas\,a$^{-1}$] & $222.67 \pm 0.02^{b}$ & $222.66 \pm 0.01$ \\
$\mu_{\delta}$ [mas\,a$^{-1}$]       & $98.13 \pm 0.02^{b}$  & $98.08 \pm 0.01$ \\
\hline
\end{tabular}
\tablefoot{
\tablefoottext{a}{From \citet{Gallenne_2023}.}
\tablefoottext{b}{From Gaia DR3 catalogue with applied correction.}
}
\end{table}

\begin{table}
\caption{Parallax comparison for the non-pulsating binary sample.}
\label{tab:plx_binary}
\centering
\small
\setlength{\tabcolsep}{4pt}
\renewcommand{\arraystretch}{1.15}
\begin{tabular}{lcccc}
\hline\hline
Name & Reference & Gaia DR3 & SS & BS \\
\hline
TZ For    & $5.37^{\pm 0.02}$     & $5.48^{\pm 0.02}$    & $5.362^{\pm 0.015}$  & $5.363^{\pm 0.022}$ \\
AK For    & $32.147^{\pm 0.055}$  & $32.19^{\pm 0.02}$   & $32.140^{\pm 0.016}$ & $32.147^{\pm 0.015}$ \\
HD 9312   & $17.56^{\pm 0.10}$    & $17.77^{\pm 0.06}$   & $17.341^{\pm 0.125}$ & $17.561^{\pm 0.038}$ \\
HD 41255  & $16.344^{\pm 0.014}$  & $16.30^{\pm 0.02}$   & $16.343^{\pm 0.021}$ & $16.344^{\pm 0.031}$ \\
HD 70937  & $15.513^{\pm 0.018}$  & $15.51^{\pm 0.03}$   & $15.510^{\pm 0.031}$ & $15.513^{\pm 0.024}$ \\
HD 210763 & $10.692^{\pm 0.037}$  & $10.66^{\pm 0.03}$   & $10.739^{\pm 0.025}$ & $10.717^{\pm 0.081}$ \\
HD 224974 & $16.176^{\pm 0.047}$  & $16.13^{\pm 0.03}$   & $16.198^{\pm 0.023}$ & $16.188^{\pm 0.047}$ \\
HD 188088 & $70.857^{\pm 0.019}$  & $70.70^{\pm 0.04}$   & $70.966^{\pm 0.025}$ & $70.903^{\pm 0.083}$ \\
LL Aqr    & $7.358^{\pm 0.020}$   & $7.33^{\pm 0.02}$    & $7.307^{\pm 0.020}$  & $7.329^{\pm 0.082}$ \\
o~Leo     & $24.412^{\pm 0.081}$  & $24.20^{\pm 0.24}$   & $24.494^{\pm 0.081}$ & $24.411^{\pm 0.078}$ \\
V963 Cen  & $8.763^{\pm 0.016}$   & $8.74^{\pm 0.02}$    & $8.765^{\pm 0.013}$  & $8.764^{\pm 0.018}$ \\
\hline
\end{tabular}
\tablefoot{
Gaia DR3 values are with the \citet{Lindegren_2021} correction applied for systems with magnitude brighter than $G=6$.
\tablefoottext{a}{TZ For values are from \citet{Gallenne_2016}. Other reference values are from \citet{Gallenne_2023}.}
}
\end{table}

\begin{table}
\caption{Proper motion corrections and corrected proper motion values for the Cepheid binaries, derived using the methodology outlined in Sect.~\ref{correction}.}
\label{tab:pm_corrections}
\centering
\small
\setlength{\tabcolsep}{4pt}
\renewcommand{\arraystretch}{1.15}
\begin{tabular}{lcccc}
\hline\hline
Name & $\Delta\mu_{\alpha^\star}$ & $\Delta\mu_{\delta}$ & $\mu_{\alpha^\star,\mathrm{corr}}$ & $\mu_{\delta,\mathrm{corr}}$ \\
     & [mas\,a$^{-1}$] & [mas\,a$^{-1}$] & [mas\,a$^{-1}$] & [mas\,a$^{-1}$] \\
\hline
V1334 Cyg & $+1.51^{\pm 0.04}$ & $+2.53^{\pm 0.05}$ & $+0.78^{\pm 0.11}$ & $-2.31^{\pm 0.13}$ \\
SU Cyg    & $+0.07^{\pm 0.05}$ & $-0.28^{\pm 0.05}$ & $+0.04^{\pm 0.09}$ & $-2.97^{\pm 0.11}$ \\
V350 Sgr  & $+0.56^{\pm 0.06}$ & $-0.73^{\pm 0.05}$ & $-1.40^{\pm 0.13}$ & $-3.50^{\pm 0.10}$ \\
U Aql     & $+1.91^{\pm 0.06}$ & $-0.62^{\pm 0.05}$ & $-0.94^{\pm 0.16}$ & $-11.35^{\pm 0.13}$ \\
S Mus     & $-0.30^{\pm 0.07}$ & $+0.35^{\pm 0.07}$ & $-7.84^{\pm 0.17}$ & $-1.86^{\pm 0.17}$ \\
AW Per    & $+0.08^{\pm 0.04}$ & $+0.85^{\pm 0.03}$ & $+0.21^{\pm 0.07}$ & $-2.2^{\pm 0.05}$ \\
\hline
\end{tabular}
\end{table}

\begin{table}
\caption{Fitted parameters of the four models for the AW~Per Gaia DR4 simulation.}
\label{tab:Aw_Per_fits}
\centering
\small
\setlength{\tabcolsep}{3pt}
\renewcommand{\arraystretch}{1.1}
\begin{tabular}{lcccc}
\hline\hline
Parameter & SS & VIMF & VIML & VIMA \\
\hline
$\chi^2_{\rm red}$ &
$3.70^{\pm0.33}$ &
$1.29^{\pm0.07}$ &
$1.13^{\pm0.07}$ &
$1.11^{\pm0.06}$ \\
\hline
$\mu_{\alpha^\star}$ &
$0.46^{\pm0.03}$ &
$0.34^{\pm0.02}$ &
$0.32^{\pm0.01}$ &
$0.34^{\pm0.02}$ \\
$\mu_\delta$ &
$-1.31^{\pm0.02}$ &
$-1.11^{\pm0.01}$ &
$-1.09^{\pm0.01}$ &
$-1.09^{\pm0.01}$ \\
$\varpi$ &
$0.88^{\pm0.04}$ &
$1.07^{\pm0.02}$ &
$1.11^{\pm0.02}$ &
$1.11^{\pm0.02}$ \\
\hline
$D_\alpha^\star$ &
-- &
$4.19^{\pm0.16}$ &
$4.01^{\pm0.18}$ &
$4.02^{\pm0.18}$ \\
$D_\delta$ &
-- &
$0.58^{\pm0.06}$ &
$0.13^{\pm0.10}$ &
$0.04^{\pm0.11}$ \\
$\dot{D}_\alpha^\star$ &
-- &
-- &
$0.22^{\pm0.09}$ &
$0.34^{\pm0.10}$ \\
$\dot{D}_\delta$ &
-- &
-- &
$-0.44^{\pm0.07}$ &
$-0.32^{\pm0.10}$ \\
$k$ &
-- &
-- &
-- &
$0.00^{\pm0.00}$ \\
$s$ &
-- &
-- &
-- &
$0.19^{\pm25.29}$ \\
\hline
$\ddot{D}_\alpha^\star$ &
-- &
-- &
-- &
$-0.001^{\pm0.006}$ \\
$\ddot{D}_\delta$ &
-- &
-- &
-- &
$-0.220^{\pm0.102}$ \\
$\dot{\mu}_{\alpha^\star}$ &
-- &
-- &
-- &
$0.00002^{\pm0.00054}$ \\
$\dot{\mu}_{\delta}$ &
-- &
-- &
-- &
$-0.0066^{\pm0.0160}$ \\
\hline
\end{tabular}
\end{table}

\begin{table}
\caption{Parallax comparison for three Cepheid systems and BH3.}
\label{tab:plx_cepheid}
\centering
\small
\setlength{\tabcolsep}{4pt}
\renewcommand{\arraystretch}{1.15}
\begin{tabular}{lcccc}
\hline\hline
Name & Reference & Gaia DR3 & SS & BS \\
\hline
SU Cyg    & $1.080^{\pm 0.006\,a}$ & $1.055^{\pm 0.052}$ & $1.063^{\pm 0.064}$ & $1.075^{\pm 0.016}$ \\
V1334 Cyg & $1.388^{\pm 0.015\,b}$ & $1.083^{\pm 0.074}$ & $1.306^{\pm 0.061}$ & $1.383^{\pm 0.023}$ \\
V350 Sgr  & $1.014^{\pm 0.049\,c}$ & $0.810^{\pm 0.062}$ & $0.839^{\pm 0.051}$ & $1.009^{\pm 0.084}$ \\
BH3       & $1.693^{\pm 0.016\,d}$ & $1.680^{\pm 0.069}$ & $1.636^{\pm 0.081}$ & $1.708^{\pm 0.017}$ \\
\hline
\end{tabular}
\tablefoot{
Gaia DR3 values are with the \citet{Lindegren_2021} correction applied for systems with magnitude brighter than six.
\tablefoottext{a}{\citet{Gallenne_2025}.}
\tablefoottext{b}{\citet{Gallenne_2018}.}
\tablefoottext{c}{\citet{Evans_2025}.}
\tablefoottext{d}{\citet{Panuzzo_2024}.}
}
\end{table}

\begin{figure*}[hbt!]
  \centering \includegraphics[width=.99\hsize]{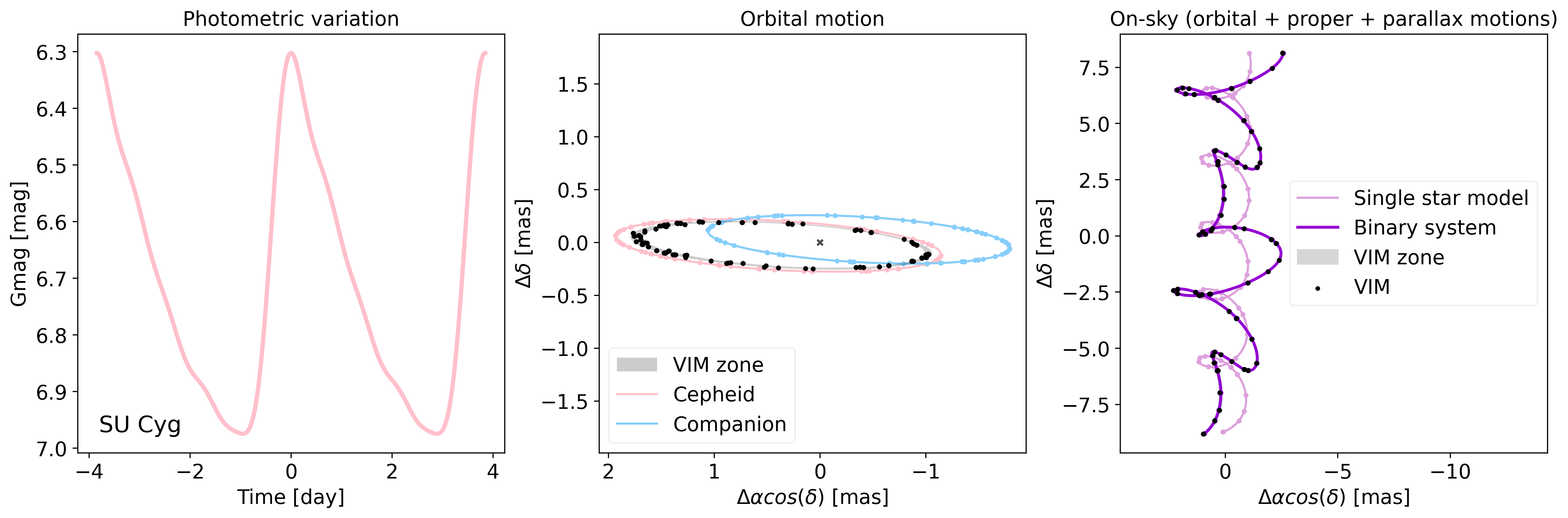}
  \caption{Figure of SU Cyg similar to V1334 Cyg (Fig.~\ref{V1334Cyg})}
  \label{SUCyg} 
\end{figure*}

\begin{figure*}[hbt!]
  \centering \includegraphics[width=.99\hsize]{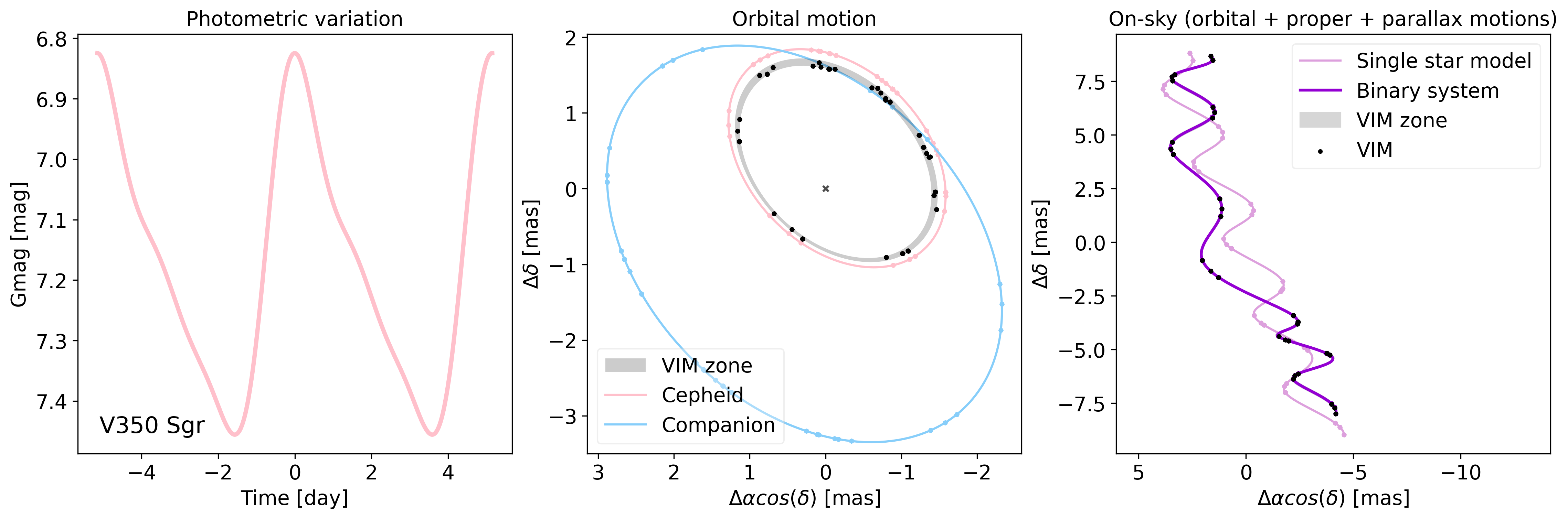}
  \caption{Figure of V350 Sgr similar to V1334 Cyg (Fig.~\ref{V1334Cyg})}
  \label{V350Sgr} 
\end{figure*}

\begin{figure*}[hbt!]
  \centering \includegraphics[width=.99\hsize]{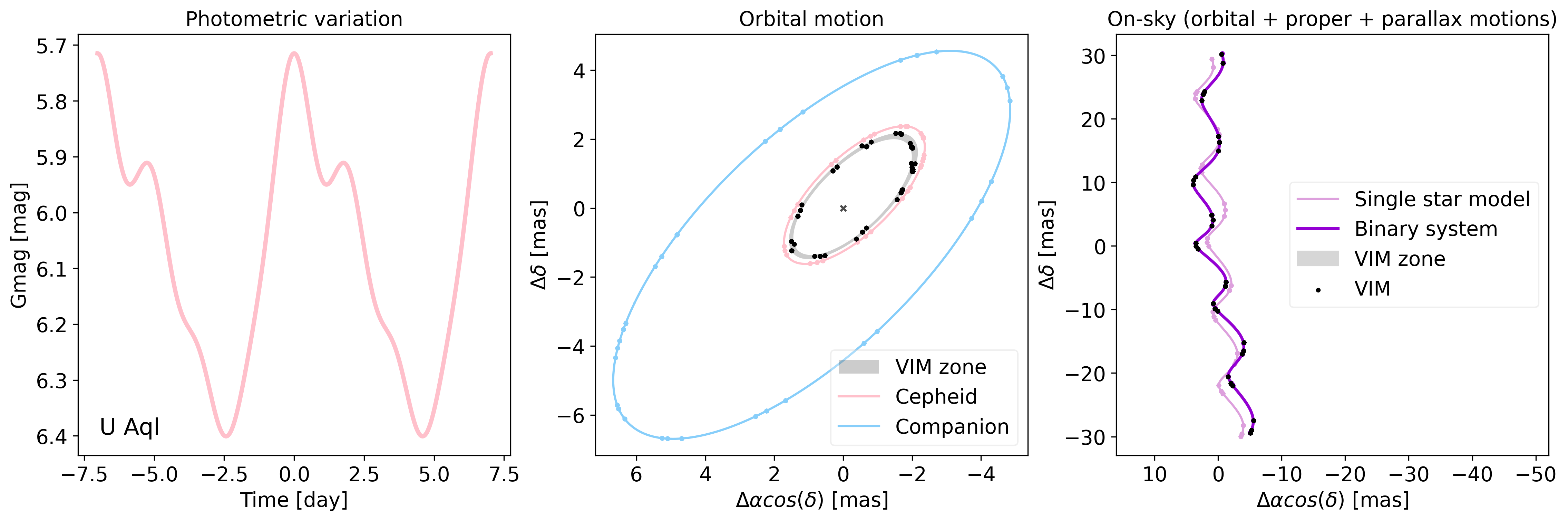}
  \caption{Figure of U Aql similar to V1334 Cyg (Fig.~\ref{V1334Cyg})}
  \label{UAql} 
\end{figure*}

\begin{figure*}[hbt!]
  \centering \includegraphics[width=.99\hsize]{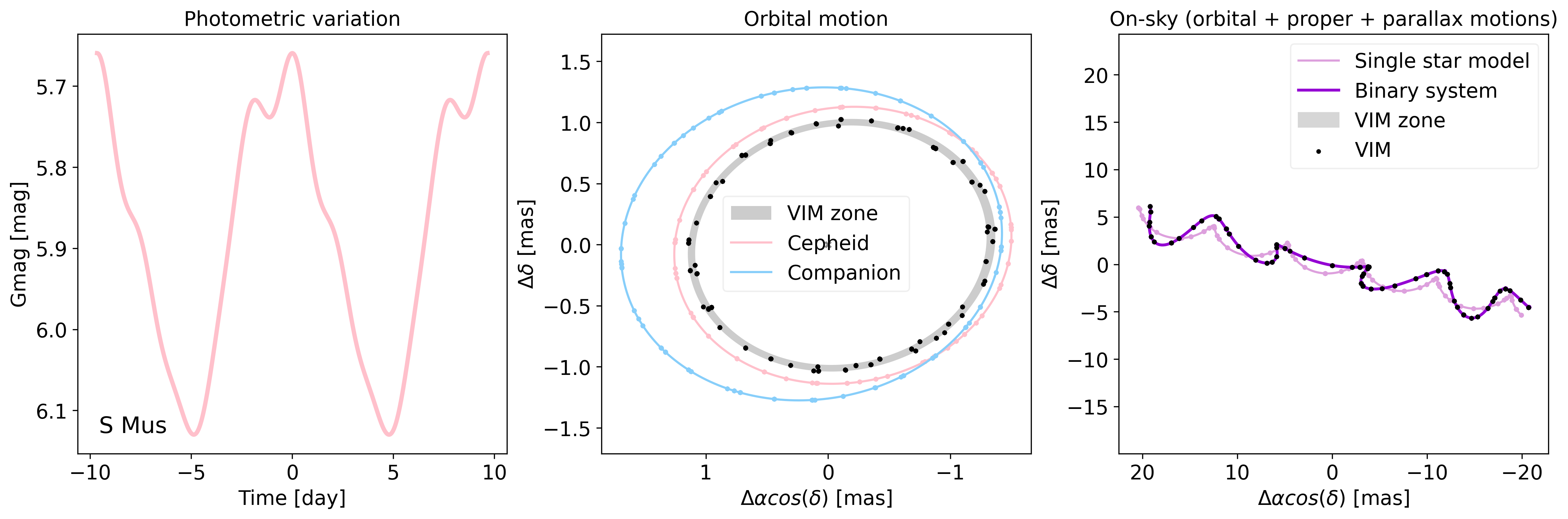}
  \caption{Figure of S Mus similar to V1334 Cyg (Fig.~\ref{V1334Cyg})}
  \label{SMus} 
\end{figure*}

\begin{figure*}[hbt!]
  \centering \includegraphics[width=.99\hsize]{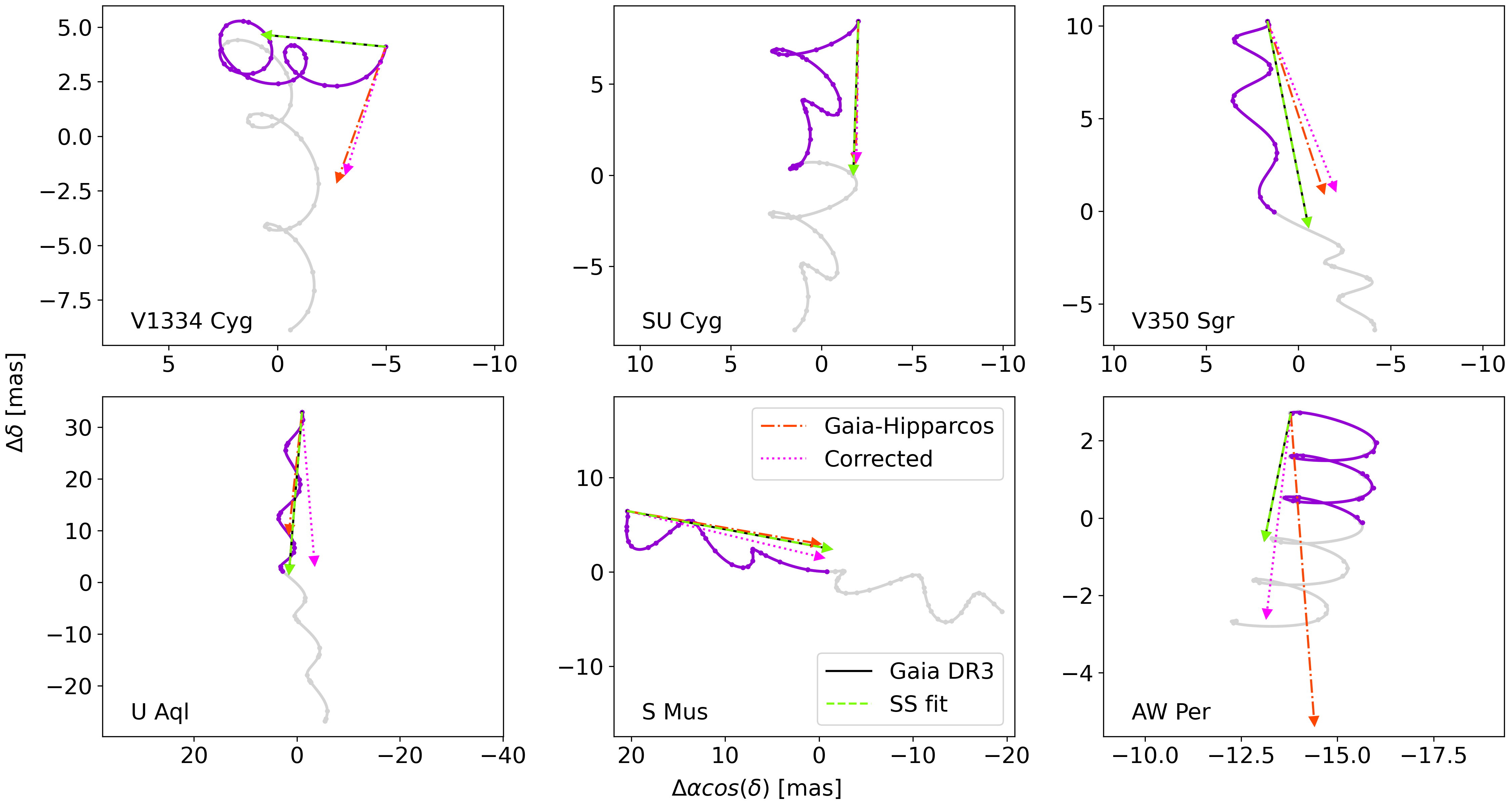}
  \caption{DR3 on-sky tracks (violet) compared with DR4 (grey) for the six Cepheids considered here. The vectors indicate the proper motion values.}
  \label{PMs_cepheids} 
\end{figure*}

\end{appendix}
\end{document}